\documentclass[letterpaper]{article} 
\usepackage{aaai2026}  
\usepackage{times}  
\usepackage{helvet}  
\usepackage{courier}  
\usepackage[hyphens]{url}  
\usepackage{graphicx} 
\usepackage{natbib}  
\usepackage{caption} 
\usepackage{algorithm}
\usepackage{algorithmic}
\usepackage[utf8]{inputenc}
\usepackage{textcomp} 
\usepackage{multirow}
\usepackage{booktabs}
\usepackage{array}
\usepackage{amsmath} 
\usepackage{amssymb}
\usepackage{amsthm}
\usepackage{xcolor}
\usepackage{colortbl}
\usepackage{amsfonts}
\usepackage{cuted}
\usepackage{dsfont}
\usepackage{placeins}
\usepackage[most]{tcolorbox}
\tcbuselibrary{breakable, listings}

\newtcblisting{promptbox}[2][]{
  enhanced,
  breakable,
  colback=blue!2!white,
  colframe=blue!60!black,
  colbacktitle=blue!60!black,
  coltitle=white,
  fonttitle=\bfseries,
  title={#2},
  boxrule=0.8pt,
  arc=2pt,
  left=4pt,
  right=4pt,
  top=4pt,
  bottom=4pt,
  listing only,
  listing options={
    basicstyle=\footnotesize\ttfamily,
    breaklines=true,
    breakatwhitespace=true,
    columns=flexible,
    keepspaces=true,
    showstringspaces=false
  },
  #1
}
\usepackage{newfloat}
\usepackage{listings}
\DeclareCaptionStyle{ruled}{labelfont=normalfont,labelsep=colon,strut=off} 
\floatstyle{ruled}
\newfloat{listing}{tb}{lst}{}
\floatname{listing}{Listing}

\title{OPERA: A Reinforcement Learning--Enhanced Orchestrated Planner-Executor Architecture for Reasoning-Oriented Multi-Hop Retrieval}

\author{
    Yu Liu\textsuperscript{\rm 1, 2},
    Yanbing Liu\textsuperscript{\rm 1, 2},
    Fangfang Yuan\textsuperscript{\rm 1},
    Cong Cao\textsuperscript{\rm 1},
    Youbang Sun\textsuperscript{\rm 4},
    Kun Peng\textsuperscript{\rm 1, 2},\\
    Weizhuo Chen\textsuperscript{\rm 1, 2},
    Jianjun Li\textsuperscript{\rm 3},
    Zhiyuan Ma\textsuperscript{\rm 3}\thanks{Corresponding author.}
}

\affiliations{
    \textsuperscript{\rm 1}Institute of Information Engineering, Chinese Academy of Sciences\\
    \textsuperscript{\rm 2}School of Cyber Security, University of Chinese Academy of Sciences\\
    \textsuperscript{\rm 3}School of Computer Science and Technology, Huazhong University of Science and Technology\\
    \textsuperscript{\rm 4}Department of Electronic Engineering, Tsinghua University\\
     
     liuyu@iie.ac.cn, mzyth@hust.edu.cn
}

\begin{document}
\renewcommand{\thefootnote}{\fnsymbol{footnote}}
\maketitle

\begin{abstract}
Recent advances in large language models (LLMs) and dense retrievers have driven significant progress in retrieval-augmented generation (RAG). However, existing approaches face significant challenges in complex reasoning-oriented multi-hop retrieval tasks: \emph{\textbf{1) Ineffective reasoning-oriented planning:}} Prior methods struggle to generate robust multi-step plans for complex queries, as rule-based decomposers perform poorly on out-of-template questions. \emph{\textbf{2) Suboptimal reasoning-driven retrieval:}} Related methods employ limited query reformulation, leading to iterative retrieval loops that often fail to locate golden documents. \emph{\textbf{3) Insufficient reasoning-guided filtering:}} Prevailing methods lack the fine-grained reasoning to effectively filter salient information from noisy results, hindering utilization of retrieved knowledge. Fundamentally, these limitations all stem from the weak coupling between retrieval and reasoning in current RAG architectures. We introduce the \textbf{O}rchestrated \textbf{P}lanner-\textbf{E}xecutor \textbf{R}easoning \textbf{A}rchitecture (\textbf{OPERA}), a novel reasoning-driven retrieval framework. OPERA's Goal Planning Module (\textbf{GPM}) decomposes questions into sub-goals, which are executed by a Reason-Execute Module (\textbf{REM}) with specialized components for precise reasoning and effective retrieval. To train OPERA, we propose Multi-Agents Progressive Group Relative Policy Optimization (\textbf{MAPGRPO}), a novel variant of GRPO. Experiments on complex multi-hop benchmarks show OPERA's superior performance, validating both the MAPGRPO method and OPERA's design. 
\end{abstract}
\begin{links}
    \link{Code}{https://github.com/Ameame1/OPERA}
    \link{Extended version}{https://arxiv.org/abs/2508.16438}
\end{links}

\begin{figure*}[t]
    \centering
    \includegraphics[width=\textwidth]{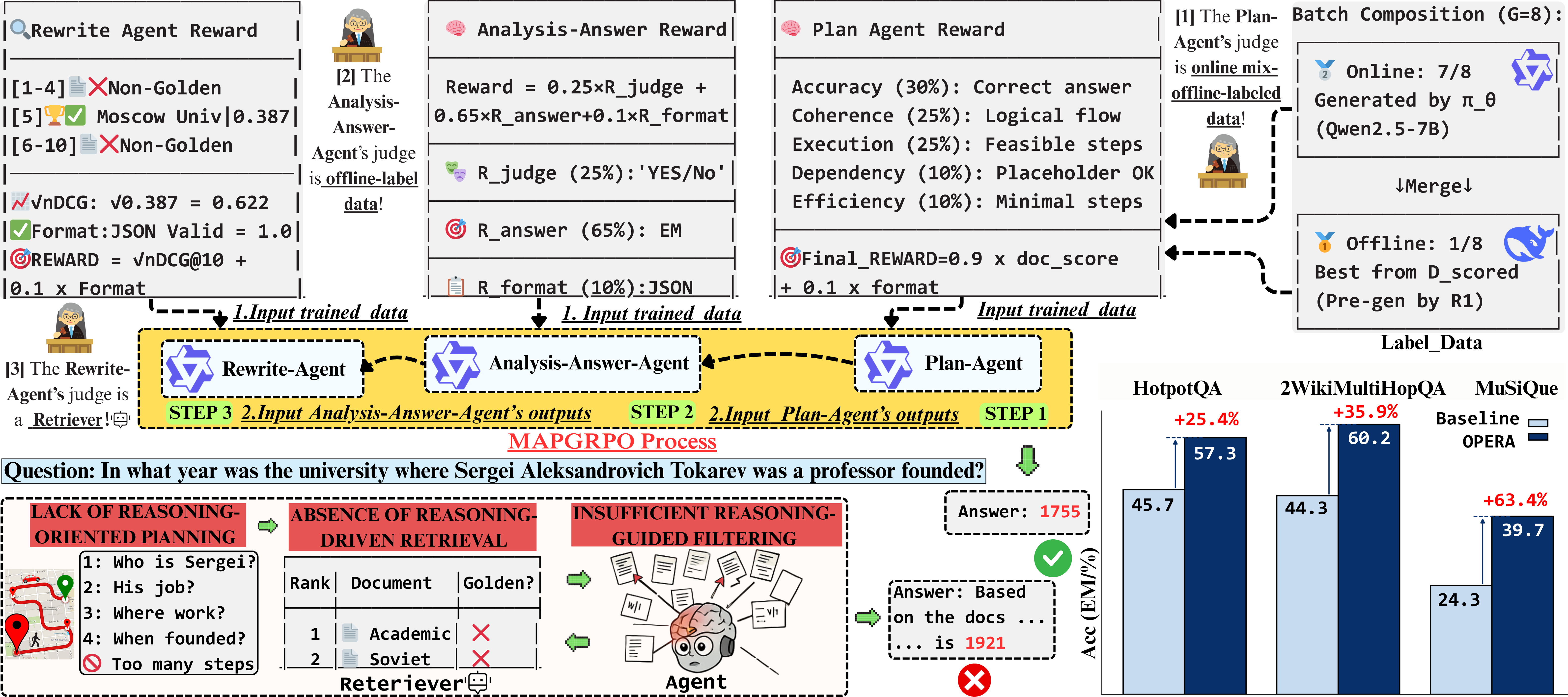}
    \caption{Overview of OPERA's MAPGRPO training framework and performance comparison with traditional RAG.}
    \label{fig:motivation}
\end{figure*}
    \section{Introduction}
    
    The ability to solve complex problems is a core aspect of intelligence, and within Retrieval-Augmented Generation (RAG), reasoning-centric retrieval provides an effective means to address these tasks in the post-training era. The concurrent improvement of Large Language Models (LLMs)~\cite{brown2020language,devlin2019bert} and dense retrieval systems~\cite{karpukhin2020dense,khattab2020colbert} has propelled the evolution of RAG. 
    The traditional RAG follows a retrieve-then-reason paradigm~\cite{lee2019latent,lewis2020retrieval}, which has now been widely optimized into a multi-stage pipeline including query rewriting, document retrieval, document filtering, and answer generation~\cite{izacard2020leveraging}. However, despite significant progress, effectively orchestrating these capabilities remains challenging, as existing approaches struggle with the demands of multi-hop reasoning.
    Consider a query such as, ``What was the previous occupation of the person who succeeded the founder of the company that acquired WhatsApp?'' Such questions demand not merely retrieving documents, but orchestrating a precise sequence of retrieval and reasoning steps where each operation depends critically on its predecessors. 
    
Current approaches face several key challenges. \textbf{First}, existing solutions have \emph{\textbf{limited reasoning-oriented planning}}. While planner-first models like PlanRAG \cite{lee2024planrag} and REAPER~\cite{reaper2024} introduce upfront planning mechanisms, their static plans cannot dynamically adapt to unforeseen challenges during retrieval. \textbf{Second}, most methods \emph{\textbf{lack effective reasoning-aware retrieval}}. Even adaptive methods like Adaptive-RAG \cite{jeong2024adaptive} and AT-RAG \cite{zhang2024atrag}, which adjust retrieval strategy based on query complexity, lack the fine-grained, reasoning-driven query reformulation needed for complex multi-hop scenarios. \textbf{Third}, current systems provide \emph{\textbf{inadequate reasoning-guided filtering}}. Even when retrieval fetches golden documents, they are often buried within noisy top-K results. While iterative approaches like ReAct~\cite{yao2022react} attempt to address this through reasoning-action loops, and recent methods such as BGM~\cite{ke2024bridging} show promise in bridging retriever-LLM preferences, their effectiveness remains limited by indirect reward signals that fail to capture the nuanced reasoning required for effective filtering. These limitations persist because advanced enhancement strategies—spanning SFT, preference optimization, and RL—are insufficient or misaligned, often leading to goal misalignment between modules~\cite{qi2023fine,wang2023instruction}. These issues stem from fundamental weakness in the coupling between retrieval and reasoning, preventing full utilization of modern LLMs and dense retrievers' capabilities.
    
    To address these limitations, we introduce OPERA, a novel reasoning-driven framework.  OPERA systematically decouples high-level strategic planning from low-level tactical execution through two core modules: a Goal Planning Module (GPM) and a Reason-Execute Module (REM).
    The GPM uses a dedicated \textbf{Plan Agent} to decompose complex queries into coherent, executable sub-goals. The REM implements a dual-agent system supported by a neural dense retriever: the \textbf{Analysis-Answer Agent} extracts precise answers from retrieved context, while a specialized \textbf{Rewrite Agent} reformulates queries to improve subsequent retrieval attempts. Furthermore, OPERA features a Trajectory Memory Component (TMC) to enhance interpretability, providing a clear rationale for each action taken by the agents.
    Our training protocol sequentially optimizes each of the three agents with a GRPO reward function tailored to its role—plan quality, reasoning accuracy, and retrieval effectiveness.
    \begin{figure*}[t]
    \centering
    \includegraphics[width=\textwidth]{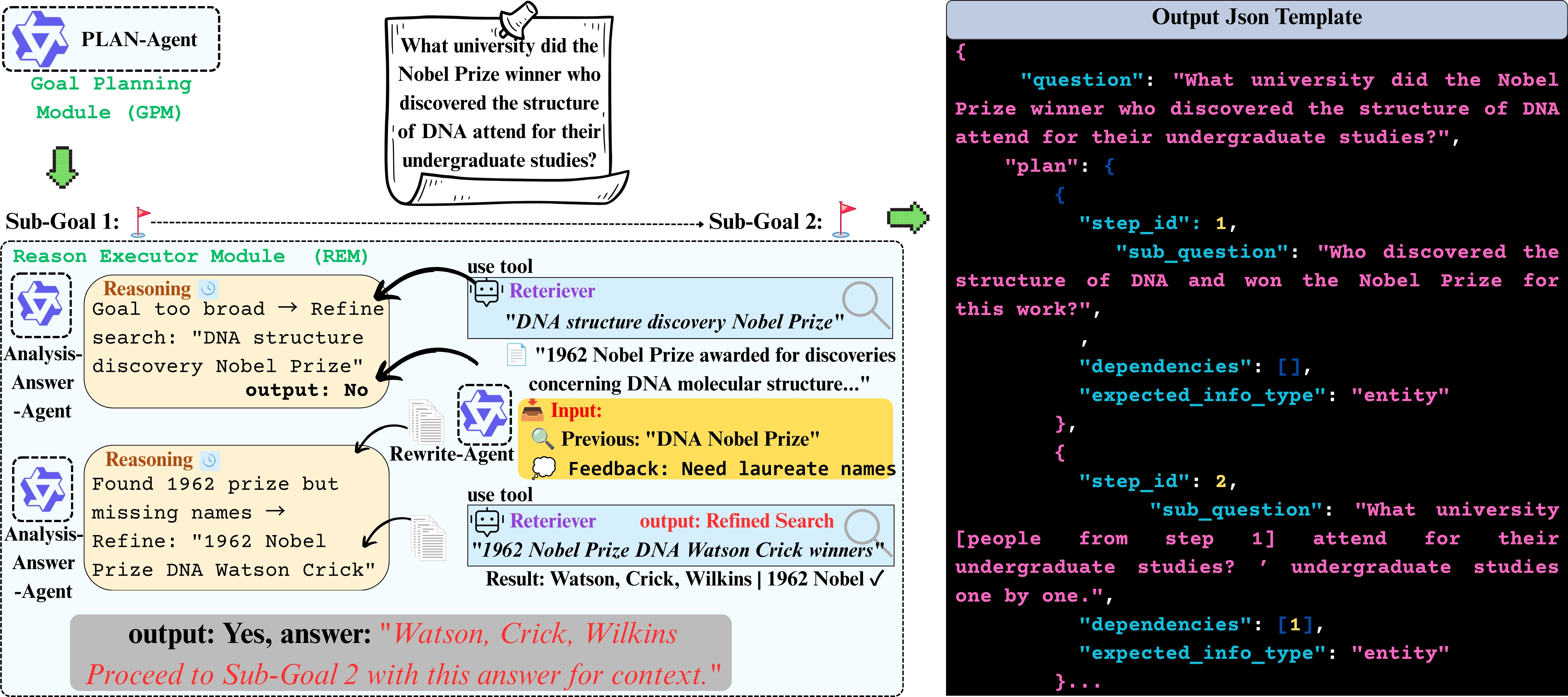}
    \caption{Overview of OPERA architecture showing the Goal Planning Module (GPM) with Plan Agent for strategic decomposition, and the Reason-Execute Module (REM) with Analysis-Answer and Rewrite Agents for adaptive execution. The Trajectory Memory Component (TMC) will record all things.}

    \label{fig:workflow}
    \end{figure*}
    Our key contributions are three-fold:
\begin{itemize}
\item \textbf{A reasoning-driven retrieval framework.} OPERA integrates reasoning into each component, improving planning, retrieval, and filtering effectiveness in RAG systems. OPERA features a TMC that enhances interpretability through action rationales.

\item \textbf{A specialized training algorithm.} MAPGRPO enhances reasoning capabilities through fine-grained, role-specific credit assignment, improving individual agent skills while ensuring coordination across the planning, retrieval, and reasoning workflow.

\item \textbf{Strong empirical results on multi-hop benchmarks.} Extensive experiments validate OPERA's reasoning-centric architecture and training approach are effective.
\end{itemize}

\section{Related Work}  
\noindent\textbf{RAG.}
To mitigate hallucination, Retrieval-Augmented Generation (RAG) was introduced to ground outputs in external knowledge \cite{lewis2020retrieval}. The initial ``retrieve-then-read'' paradigm~\cite{lee2019latent}, however, proved insufficient for multi-hop reasoning due to its static pipeline~\cite{trivedi2022musique,arabzadeh2021predicting,luan2021sparse}. The field has since evolved toward more dynamic retrieval, for instance by routing queries based on complexity \cite{jeong2024adaptive} or introducing explicit planning \cite{lee2024planrag}. While these methods improve upon static RAG, they primarily optimize the retrieval act itself, rather than the overarching reasoning strategy. OPERA is distinct in its hierarchical approach: a specialized Goal Planning Module (GPM) governs high-level strategy, while an agentic Reason-Execute Module (REM) handles tactical execution, including fine-grained analysis and adaptive query reformulation.
    
\noindent\textbf{Chain-of-Thought.}
Chain-of-Thought (CoT) methods \cite{wei2022chain,yao2023tree} and agentic frameworks like ReAct \cite{yao2022react} and IRCoT \cite{trivedi2023interleaving} established the importance of decomposing problems and interleaving reasoning with actions. However, these frameworks rely on a single, general LLM for high-level planning and low-level execution, which can compromise reliability. While multi-agent systems like MetaGPT \cite{hong2024metagpt} assign distinct roles, they typically use generalist models. OPERA advances by employing an asymmetric architecture, pairing the strategic Goal Planning Module (GPM) with the agentic Reason-Execute Module (REM) to separate strategic and tactical concerns.

\noindent\textbf{Reinforcement Learning.}
A key aspect of our work is the training of our specialized planner and executor. Many frameworks utilize on-policy algorithms like Proximal Policy Optimization (PPO); however, PPO often struggles with the large action spaces and sparse rewards common in RAG~\cite{stiennon2020learning,fernandezoLooria2022survey,ramamurthy2023reinforcement}, making training unstable and sample-inefficient. To address this, preference-based optimization has gained traction. While Direct Preference Optimization (DPO) \cite{rafailov2023direct} has become a standard for learning from binary preferences (chosen $>$ rejected), it is ill-suited for more nuanced reward signals~\cite{ivison2024unpacking}. Our training methodology generates fine-grained scalar scores reflecting the quality of a plan or an execution step. Using DPO would discard this rich information by compressing it into a binary signal. To fully leverage these scalar rewards within our multi-agent setting, we build upon Group Relative Policy Optimization (GRPO)~\cite{shao2024grpo} and introduce our own variant: Multi-Agents Progressive Group Relative Policy Optimization (MAPGRPO). Unlike standard GRPO, MAPGRPO is specifically designed for our staged training protocol, enabling fine-grained credit assignment~\cite{papangelis2019collaborative} and ensuring coordinated optimization across the distinct roles of the GPM and REM agents.
    
    \section{Method}
    \subsection{Problem Formulation}
    
    We formalize reasoning-driven multi-hop retrieval as follows. Given a complex question $q$, the goal is to generate an accurate answer $a^*$ through orchestrated reasoning and retrieval operations. Let $\mathcal{D}$ denote the document corpus and $\mathcal{R}: \mathcal{Q} \rightarrow \mathcal{D}^k$ be a retrieval function that maps queries from query space $\mathcal{Q}$ to top-$k$ documents.

    The task decomposes into three reasoning-driven sub-problems: (1) \textbf{Reasoning-Driven Planning}: $f_{\text{plan}}: q \rightarrow \{p_1, ..., p_m\}$ where $m$ represents the number of sub-goals, (2) \textbf{Reasoning-Driven Retrieval}: $f_{\text{rewrite}}: (p_i, \mathcal{D}_i^{\text{insuf}}) \rightarrow q'_i$ where $\mathcal{D}_i^{\text{insuf}}$ denotes insufficient documents and $q'_i$ is the reformulated query, and (3) \textbf{Reasoning-Driven Answering}: $f_{\text{exec}}: (p_i, \mathcal{D}_i) \rightarrow (a_i, \phi_i)$ where $a_i$ is the answer and $\phi_i \in \{0,1\}$ guides conditional execution.

    \subsection{Overview and Architecture}
    We introduce OPERA (Orchestrated Planner-Executor Reasoning Architecture), a framework that systematically decouples strategic planning from tactical execution. As illustrated in Figures \ref{fig:motivation} and \ref{fig:workflow}, OPERA operates through two core modules: the \textbf{Goal Planning Module (GPM)} containing a Plan Agent for strategic decomposition, and the \textbf{Reason-Execute Module (REM)} containing Analysis-Answer and Rewrite Agents for conditional execution and adaptive retrieval. The Plan Agent decomposes complex questions into sub-goals $\mathcal{P}$ (plan consisting of sub-goals) with placeholder dependencies. The Analysis-Answer Agent performs information sufficiency assessment $\phi$ (information sufficiency indicator) and answer extraction from retrieved documents $\mathcal{D}_i$ (documents retrieved for sub-goal $i$). The Rewrite Agent reformulates queries when information is insufficient. To optimize this multi-agent system, we introduce MAPGRPO for sequential training with role-specific rewards.

\begin{algorithm}[t]
\caption{MAPGRPO Training for OPERA}
\label{alg:mapgrpo}
\small
\begin{algorithmic}[1]
\REQUIRE Dataset $\mathcal{D}$, group size $G$, KL coefficient $\beta$, pre-scored dataset $\mathcal{D}_{\text{scored}}$
\ENSURE Optimized parameters $\{\theta_{\text{plan}}^*, \theta_{\text{ana}}^*, \theta_{\text{rew}}^*\}$

\STATE \textbf{Stage 1: Plan Agent Training}
\FOR{epoch $e = 1$ to $E_1$}
    \FOR{batch $(q, \mathcal{G}) \in \mathcal{D}$}
        \STATE $\mathcal{C}_{\text{plan}} \leftarrow \{c_1, ..., c_{G-1}\} \sim \pi_{\text{plan}}^{(\theta)}(\cdot|q)$ \COMMENT{Generate $G-1$ candidates}
        \STATE $c_{\text{best}} \leftarrow \arg\max_{c \in \mathcal{D}_{\text{scored}}(q)} r_{\text{pre}}(q, c)$ \COMMENT{Select best pre-scored sample}
        \STATE $\mathcal{C}_{\text{plan}} \leftarrow \mathcal{C}_{\text{plan}} \cup \{c_{\text{best}}\}$ \COMMENT{Add to candidate set}
        \STATE Compute rewards $\{r_{\text{plan}}(q, c)\}_{c \in \mathcal{C}_{\text{plan}}}$
        \STATE Update $\theta_{\text{plan}}$ via GRPO loss with advantages from Eq. (2)
    \ENDFOR
\ENDFOR

\STATE \textbf{Stage 2: Analysis-Answer Agent Training}
\FOR{epoch $e = 1$ to $E_2$}
    \FOR{batch $(p, \mathcal{D}, a^*) \in \mathcal{D}_{\text{exec}}(\theta_{\text{plan}}^*)$}
        \STATE $\mathcal{C}_{\text{ana}} \leftarrow \{c_1, ..., c_{G-1}\} \sim \pi_{\text{ana}}^{(\theta)}(\cdot|p, \mathcal{D})$
        \STATE $c_{\text{best}} \leftarrow \arg\max_{c \in \mathcal{D}_{\text{scored}}(p)} r_{\text{pre}}(p, \mathcal{D}, c)$
        \STATE $\mathcal{C}_{\text{ana}} \leftarrow \mathcal{C}_{\text{ana}} \cup \{c_{\text{best}}\}$
        \STATE Compute rewards $\{r_{\text{ana}}(p, \mathcal{D}, c)\}_{c \in \mathcal{C}_{\text{ana}}}$
        \STATE Update $\theta_{\text{ana}}$ via GRPO loss
    \ENDFOR
\ENDFOR

\STATE \textbf{Stage 3: Rewrite Agent Training}
\FOR{epoch $e = 1$ to $E_3$}
    \FOR{batch $(p, \mathcal{D}_{\text{insuf}}, \mathcal{G}) \in \mathcal{D}_{\text{neg}}$}
        \STATE $\mathcal{C}_{\text{rew}} \leftarrow \{c_1, ..., c_{G-1}\} \sim \pi_{\text{rew}}^{(\theta)}(\cdot|p, \mathcal{D}_{\text{insuf}})$
        \STATE $c_{\text{best}} \leftarrow \arg\max_{c \in \mathcal{D}_{\text{scored}}(p)} r_{\text{pre}}(p, c)$
        \STATE $\mathcal{C}_{\text{rew}} \leftarrow \mathcal{C}_{\text{rew}} \cup \{c_{\text{best}}\}$
        \STATE Compute rewards $\{r_{\text{rew}}(p, c)\}_{c \in \mathcal{C}_{\text{rew}}}$
        \STATE Update $\theta_{\text{rew}}$ via GRPO loss
    \ENDFOR
\ENDFOR

\RETURN $\{\theta_{\text{plan}}^*, \theta_{\text{ana}}^*, \theta_{\text{rew}}^*\}$
\end{algorithmic}
\end{algorithm}

    \subsection{Multi-Agents Progressive Group Relative Policy Optimization (MAPGRPO)}
    We propose MAPGRPO, a novel variant of Group Relative Policy Optimization (GRPO) \cite{shao2024grpo}.
    
    \noindent \textbf{Theoretical Foundation.} Given a policy $\pi_\theta$ parameterized by $\theta$, GRPO optimizes objective:
    
    \begin{equation}
    \mathcal{J}_{\text{GRPO}}(\theta) = \mathbb{E}_{x \sim \mathcal{D}} \left[ \mathbb{E}_{y_i \sim \pi_\theta(\cdot|x)} \left[ A_i(x, y_i) \right] \right],
    \end{equation}
    where $x$ is the input, $y_i$ denotes the $i$-th generated output, and the advantage function $A_i$ is computed relative to the group mean:
    \begin{equation}
    A_i(x, y_i) = r(x, y_i) - {\textstyle \frac{1}{G} \sum_{j=1}^G r(x, y_j)}.
    \end{equation}
    Here, $G$ (group size) is the number of candidates in each group, $r(x, y_i)$ is the reward for the $i$-th sample, and $\bar{r}(x)$ serves as a baseline computed from the current batch. The policy gradient is then:
    \begin{equation}
    \nabla_\theta \mathcal{J}_{\text{GRPO}} = \mathbb{E} \left[ \sum_{i=1}^G A_i(x, y_i) \nabla_\theta \log \pi_\theta(y_i|x) \right].
    \end{equation}
    To prevent policy collapse, GRPO incorporates a KL divergence constraint with coefficient $\beta$ (KL divergence coefficient controlling the strength of regularization):
    \begin{equation}
    \mathcal{L}_{\text{GRPO}}(\theta) = -\mathcal{J}_{\text{GRPO}}(\theta) + \beta \mathbb{D}_{\text{KL}}[\pi_\theta || \pi_{\text{ref}}],
    \end{equation}
    where $\pi_{\text{ref}}$ denotes the reference policy (typically the initial model) and $\mathbb{D}_{\text{KL}}$ represents the Kullback-Leibler divergence.
    
    \noindent\textbf{Definition 1: MAPGRPO.} Given $N$ specialized agents $\{\pi^{(k)}\}_{k=1}^N$ with heterogeneous reward functions $\{r^{(k)}\}_{k=1}^N$, MAPGRPO optimizes each agent sequentially:
    \begin{equation}
    \theta_k^* = \arg\max_{\theta_k} \mathcal{J}_{k}(\theta_k | \theta_{<k}^*),
    \end{equation}
    where $\theta_{<k}^* = \{\theta_1^*, ..., \theta_{k-1}^*\}$ represents the parameters of previously optimized agents, and:
    \begin{equation}
    \resizebox{0.90\columnwidth}{!}{$\displaystyle
        \mathcal{J}_k(\theta_k | \theta_{<k}^*) = \mathbb{E}_{x \sim \mathcal{D}_k(\theta_{<k}^*)} \left[ \mathbb{E}_{y_i \sim \pi_k^{(\theta_k)}} \left[ A_i^{(k)}(x, y_i) \right] \right].
    $}
    \end{equation}
    
    Here, $\mathcal{D}_k(\theta_{<k}^*)$ represents the distribution induced by previously trained agents, ensuring each agent adapts to its actual execution environment.
    
    \noindent\textbf{Principle.} MAPGRPO differs from standard GRPO in several key ways. First, it uses heterogeneous reward functions for specialized agents instead of homogeneous objectives. Second, it employs sequential optimization to address credit assignment problems in multi-agent training. Third, each agent trains on distributions induced by its predecessors, providing realistic execution conditions.

    \noindent\textbf{Plan Agent.} The Plan agent $\pi_{\text{plan}}$ decomposes queries into sub-goals with placeholder dependencies. Given query $q$, it generates a plan $\mathcal{P} = \{p_1, ..., p_m\}$ where each $p_i$ may contain placeholders $[t \text{ from step } j]$ for $j < i$, with $t$ indicating the expected information type (entity, location, etc.) and $j$ (step index in placeholder) the dependency step.
    
    The \textbf{reward function} is:
\begin{equation}
\begin{split}
    r_{\text{plan}}(q, \mathcal{P}) ={}& \lambda_1 \cdot f_{\text{logic}}(q, \mathcal{P}) + \lambda_2 \cdot f_{\text{struct}}(\mathcal{P}) \\
                                        & + \lambda_3 \cdot f_{\text{exec}}(\mathcal{P}, \mathcal{E}).
\end{split}
\end{equation}
where $f_{\text{logic}}$ measures decomposition validity, $f_{\text{struct}}$ evaluates placeholder syntax correctness, $f_{\text{exec}}$ represents end-to-end execution success, and $\lambda_1, \lambda_2, \lambda_3$ are weighting coefficients with $\lambda_1 + \lambda_2 + \lambda_3 = 1$.
  
    \noindent\textbf{Analysis-Answer Agent.}
    The Analysis-Answer agent $\pi_{\text{ana}}$ performs information sufficiency assessment and answer extraction. For sub-goal $p_i$ and documents $\mathcal{D}_i$:
    
    \begin{equation}
    \pi_{\text{ana}}(p_i, \mathcal{D}_i) = \begin{cases}
    (y_i, a_i, c_i) & \text{if } \phi(p_i, \mathcal{D}_i) = 1 \\
    (n_i, \perp, \rho_i) & \text{if } \phi(p_i, \mathcal{D}_i) = 0
    \end{cases},
    \end{equation}
    
    where $\phi$ is the sufficiency indicator function, $y_i$ denotes YES decision output, $a_i$ is the extracted answer, $c_i$ is confidence score, $n_i$ denotes NO decision output, $\perp$ represents null answer, and $\rho_i$ represents missing information type.
    
    The \textbf{reward function} is:
    
    \begin{equation}
    \begin{split}
        r_{\text{ana}}(p_i, \mathcal{D}_i, o_i) ={}& \alpha \cdot \mathbb{I}[\phi = \phi^*] + \beta \cdot \text{EM}(a_i, a_i^*) \\
                                            & + \gamma \cdot f_{\text{format}}(o_i),
    \end{split}
    \end{equation}
   \begin{table*}[]
\centering
\footnotesize
\renewcommand{\arraystretch}{0.85}
\begin{tabular}{l|cc|cc|cc|c}
\toprule
\multirow{2}{*}{\textbf{Method}} & \multicolumn{2}{c|}{\textbf{HotpotQA}} & \multicolumn{2}{c|}{\textbf{2WikiMultiHopQA}} & \multicolumn{2}{c|}{\textbf{Musique}} & \multirow{2}{*}{\textbf{Type}} \\
\cmidrule{2-7}
 & EM (\%) & F1 (\%) & EM (\%) & F1 (\%) & EM (\%) & F1 (\%) & \\
\midrule
Qwen2.5-7B (No Retrieval) & 18.5 & 26.8 & 16.2 & 23.7 & 4.1 & 9.1 & Single LLM \\
Single-Step RAG & 31.5 & 44.2 & 25.9 & 37.6 & 14.1 & 18.4 & Naive RAG \\
IRCoT~\cite{trivedi2023interleaving} & 42.7 & 54.8 & 43.3 & \underline{56.2} & 18.8 & 23.9 & CoT \\
OPERA (CoT) & 44.9 & \underline{58.5} & 42.3 & 50.7 & 21.2 & 32.1 & CoT \\
Adaptive-RAG \cite{jeong2024adaptive} & \underline{45.7} & 56.9 & 30.1 & 39.3 & \underline{24.3} & \underline{35.7} & SFT \\
BGM~\cite{ke2024bridging} & 41.5 & 53.8 & \underline{44.3} & 55.8 & 19.6 & 26.8 & RL \\
\textbf{OPERA (MAPGRPO)} & \textbf{57.3}{\tiny\textbf{(+11.6)}} & \textbf{69.5}{\tiny\textbf{(+11.0)}} & \textbf{60.2}{\tiny\textbf{(+15.9)}} & \textbf{72.7}{\tiny\textbf{(+16.5)}} & \textbf{39.7}{\tiny\textbf{(+15.4)}} & \textbf{58.0}{\tiny\textbf{(+22.3)}} & {RL} \\
\bottomrule
\end{tabular}
\begin{flushleft}
\textsuperscript{a}All SFT and RL methods are trained on mixed datasets (Musique+HotpotQA+2WikiMultiHopQA). Numbers in parentheses show improvement over best baseline.
\end{flushleft}
\caption{Main experimental results on three multi-hop QA benchmarks (underlined: best baseline).\textsuperscript{a}}
\label{tab:main_results}
\end{table*}        
    where $o_i$ is the output tuple, $\mathbb{I}[\cdot]$ is the indicator function, $\phi^*$ is the ground-truth sufficiency, EM denotes exact match score, $a_i^*$ is the ground-truth answer, and weights $\alpha, \beta, \gamma$ satisfy $\alpha + \beta + \gamma = 1$.

    \noindent\textbf{Rewrite Agent.} The Rewrite agent $\pi_{\text{rew}}$ reformulates queries when Analysis-Answer agent determines insufficient information. The \textbf{reward function} combines retrieval effectiveness and format compliance:
    
    \begin{equation}
    \begin{split}
        r_{\text{rew}}(q, q', \mathcal{R}) ={}& \omega_1 \cdot \sqrt{\text{NDCG@}k(\mathcal{R}(q'), \mathcal{G})} \\
                                            & + \omega_2 \cdot f_{\text{format}}(q').
    \end{split}
    \end{equation}

    where $q'$ is the rewritten query, $\mathcal{R}(q')$ represents documents retrieved using $q'$, $\mathcal{G}$ denotes golden documents, NDCG@$k$ is the normalized discounted cumulative gain at rank $k$, $\text{Score}_{\text{format}}$ evaluates query format quality, and weights $\omega_1, \omega_2$ satisfy $\omega_1 + \omega_2 = 1$ with $\omega_1 \gg \omega_2$ to prioritize retrieval effectiveness.
    
    \noindent \textbf{High-Score Sample Selection Strategy.} To address reward sparsity in early training, we select high-scoring samples from pre-scored offline data into each candidate group. For a training instance with query $q$, we generate candidates $\mathcal{C}$ (set of candidates) = $\{c_1, ..., c_{G-1}, c_{\text{best}}\}$ where:
    \begin{equation}
    c_{\text{best}} = \arg\max_{c \in \mathcal{D}_{\text{scored}}} r_{\text{pre}}(q, c).
    \end{equation}
The best candidate is selected from pre-scored dataset $\mathcal{D}_{\text{scored}}$, which contains samples generated by \textbf{large-scale} LLMs and scored through end-to-end execution, ensuring at least one high-reward sample per golden plan group. Selection ratio is maintained at 1/$G$ throughout training.  This means selecting the best candidate golden paln from multiple outputs.
    
    \subsection{Theoretical Analysis}
    
We provide rigorous theoretical foundations for OPERA's design choices. Our analysis establishes that MAPGRPO converges to local optima with rate $\mathcal{O}(1/\sqrt{T})$ under standard regularity conditions, while our reward functions are information-theoretically optimal by maximizing respective components of the mutual information decomposition $I(Q; A|D)$. Furthermore, our three-agent architecture achieves computational complexity $\mathcal{O}(h \cdot s \cdot r)$ compared to exponential $\mathcal{O}(s^h \cdot r^h)$ scaling for single-agent approaches, and our high-score sample selection strategy reduces the stochastic variance of the group reward baseline by incorporating one deterministic high-score reference per group, accelerating convergence while maintaining exploration diversity. Detailed proofs and formal statements are provided in Appendix A.5.

\section{Experimental}
    \subsection{Experimental Setup}
    More training and experiment details are in Appendix A.2.
\noindent\textbf{Datasets.} We evaluate OPERA on three multi-hop reasoning benchmarks: \textbf{HotpotQA}~\cite{yang2018hotpotqa} (90K questions), \textbf{2WikiMultiHopQA}~\cite{ho2020constructing} (150K questions), and \textbf{Musique}~\cite{trivedi2022musique} (25K questions). For out-of-domain evaluation, we use \textbf{NQ}~\cite{kwiatkowski2019natural} and \textbf{MultiHopRAG}~\cite{tang2024multihop}.

\noindent\textbf{Implementation Details.} We use Qwen2.5-7B-Instruct~\cite{yang2024qwen2.5} for Plan and Analysis-Answer agents and other baseline's backbone, Qwen2.5-3B-Instruct~\cite{yang2024qwen2.5} for the Rewrite agent, and BGE-M3~\cite{chen2024bge} as our dense retriever with top-5 document retrieval. For pre-scored dataset construction, we utilize the DeepSeek R1~\cite{deepseek2025r1} API as the data generation model.

\subsection{Main Result}
\noindent\textbf{Baselines.} We compare OPERA against methods in four categories: \textbf{Naive}: (1) Qwen2.5-7B (No Retrieval); (2) Single-Step RAG. \textbf{CoT}: (3) IRCoT~\cite{trivedi2023interleaving}; (4) OPERA (CoT Only). \textbf{SFT}: (5) Adaptive-RAG~\cite{jeong2024adaptive}. \textbf{RL}: (6) BGM~\cite{ke2024bridging}; (7) OPERA (MAPGRPO). All SFT and RL methods are trained on mixed datasets, with Plan Agent utilizing pre-scored dataset for high-score sample selection. Baselines use defaults with Faiss~\cite{douze2024faiss}.
    
    \noindent\textbf{Evaluation Metrics.} We report \textbf{EM (\%)} (exact match), \textbf{F1 (\%)} (token-level overlap), \textbf{Steps} (average reasoning steps), \textbf{Latency} (processing time), and \textbf{Success Rate} (execution completion rate).
Table~\ref{tab:main_results} shows OPERA's performance across all benchmarks.

\noindent\textbf{Performance Scales with Difficulty.} OPERA shows larger improvements on more challenging datasets—63.4\% relative improvement on Musique (from 24.3\% to 39.7\% EM) versus 25.4\% relative improvement on HotpotQA (from 45.7\% to 57.3\% EM). This suggests our approach works better for complex multi-hop reasoning tasks.

\noindent\textbf{Comparison with RL Methods.} BGM applies RL to bridge retriever-LLM gaps but achieves only 19.6\% EM on Musique. OPERA reaches 39.7\% EM on the same dataset, indicating that specialized agent architecture provides benefits beyond RL optimization alone.

\begin{table}[t!]
\centering
\begin{tabular}{lcc}
\toprule
\textbf{Configuration} & \textbf{EM (\%)} & \textbf{F1 (\%)} \\
\midrule
\multicolumn{3}{l}{\textit{Module Ablation}} \\
\midrule
w/o Plan Agent & 17.1{\tiny{(-22.6)}} & 28.5{\tiny{(-29.5)}} \\
w/o Rewrite Agent & 34.5{\tiny{(-5.2)}} & 51.8{\tiny{(-6.2)}} \\
w/o Plan \& Rewrite & 16.7{\tiny{(-23.0)}} & 27.2{\tiny{(-30.8)}} \\
\midrule
\multicolumn{3}{l}{\textit{Training Ablation}} \\
\midrule
CoT & 21.2{\tiny{(-18.5)}} & 32.1{\tiny{(-25.9)}} \\
SFT & 24.3{\tiny{(-15.4)}} & 38.2{\tiny{(-19.8)}} \\
GRPO & 34.8{\tiny{(-4.9)}} & 51.5{\tiny{(-6.5)}} \\
\midrule
\textbf{OPERA (MAPGRPO)} & \textbf{39.7} & \textbf{58.0} \\
\bottomrule
\end{tabular}
\begin{flushleft}
\footnotesize
\end{flushleft}
\caption{Module and Training Ablation}
\label{tab:combined_ablation}
\end{table}

\subsection{Ablation Studies}
  We select MuSiQue, the most challenging dataset from our main results, for ablation experiments. All training
  method variants (CoT, SFT, GRPO) use the OPERA architecture but differ in their optimization approach, and are
  trained on decomposed sub-problems—(sub-question, documents, sub-answer) tuples—to isolate training methodology
   impact from architectural contributions.

  \noindent\textbf{Architecture Has Larger Impact Than Training.}
  Table~\ref{tab:combined_ablation} shows that removing architectural components causes catastrophic performance
  drops, while training method improvements are more gradual. Removing the Plan Agent reduces performance from
  39.7\% to 17.1\% EM—below even the untrained CoT baseline (21.2\% EM)—as retrieval and reasoning modules
  receive poorly formed queries, leading to cascading errors. The Rewrite Agent has smaller but crucial impact
  (reducing EM to 34.5\%): while many questions succeed through direct retrieval, it converts otherwise failed
  cases into successful retrievals. Most strikingly, removing both components simultaneously drops performance to
   16.7\% EM—worse than removing either alone—indicating that OPERA's components form an integrated system where
  each module depends on others functioning properly.

  \noindent\textbf{Training Methods Show Clear Progression.}
  The progression across training methods follows distinct patterns. SFT improves over CoT from 21.2\% to 24.3\%
  EM through pattern learning. The jump to GRPO (34.8\% EM) comes from trajectory-level optimization, where the
  model learns effective reasoning paths rather than just correct answers. MAPGRPO's improvement to 39.7\% EM
  shows that specialized reward functions and sequential training better match the distinct requirements of
  planning, reasoning, and retrieval. However, even optimal training cannot compensate for missing architectural
  components—the gap between full OPERA and ablated versions persists across all training methods.

  These results indicate that OPERA's performance gains stem from the synergistic blend of architectural design and training: while specialized agents with defined responsibilities provide the foundational framework, coordinated optimization through MAPGRPO ensures effective collaboration in planning, retrieval, and reasoning workflow.


\begin{figure}[ht!]
\centering
\includegraphics[width=0.46\textwidth]{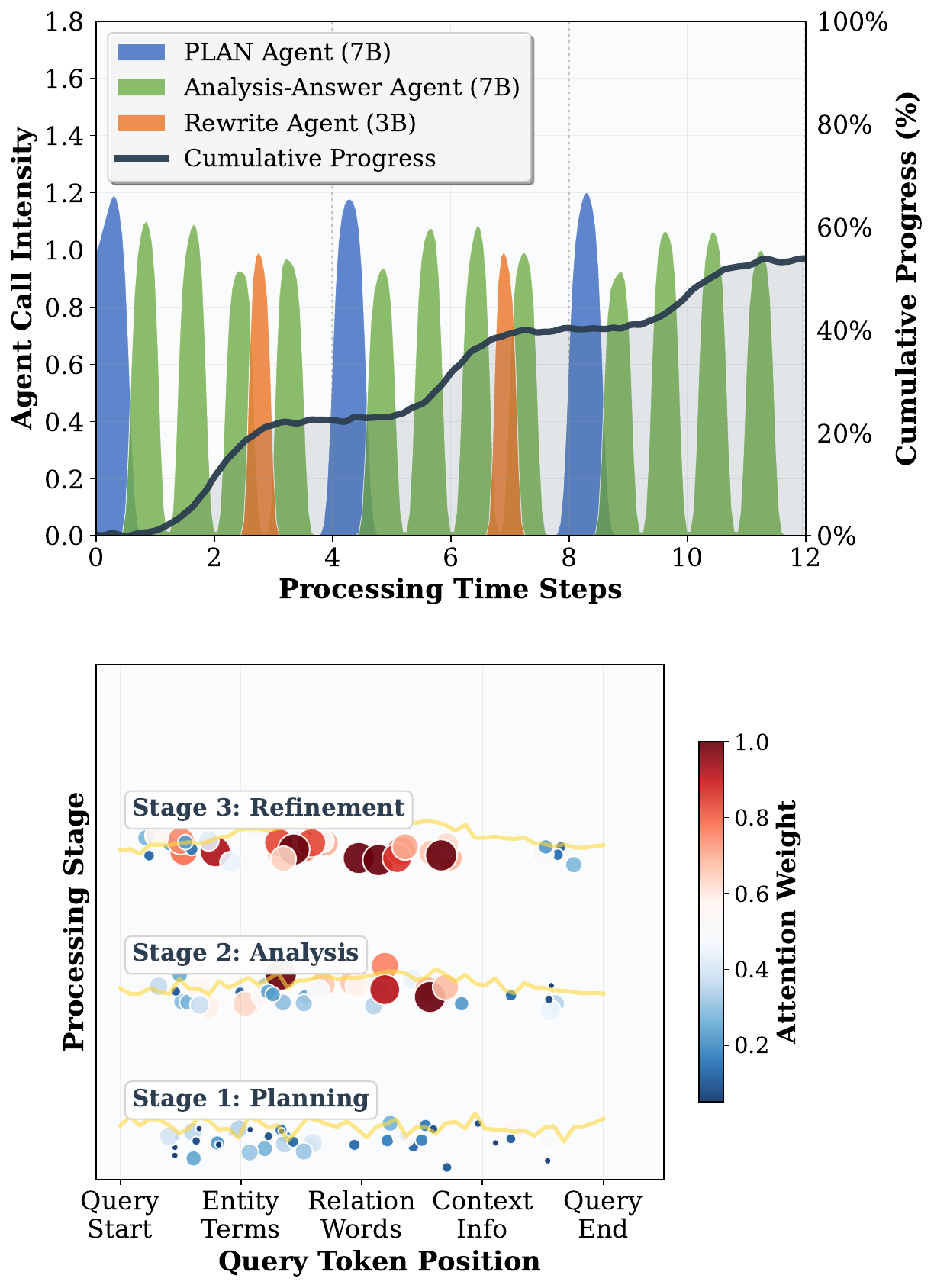}
\caption{OPERA's runtime dynamics. (\textbf{Top}) Agent call intensity and question completion rate over processing steps. (\textbf{Bottom}) Attention visualization across query token types and processing stages.}
\label{fig:runtime_dynamics}
\end{figure} 
    \begin{figure}[t]
\centering
\includegraphics[width=0.48\textwidth]{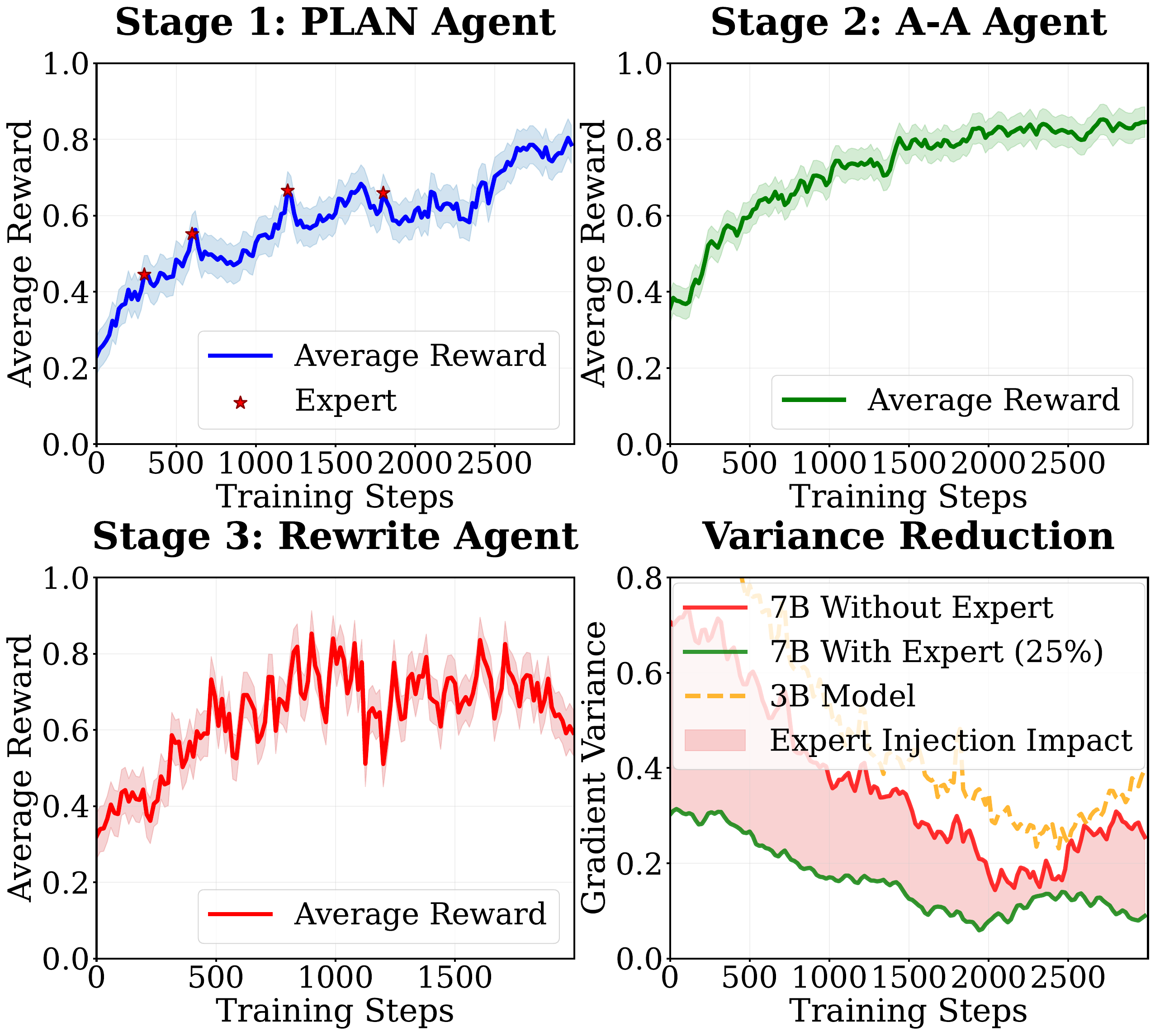}
\caption{Training dynamics for MAPGRPO across three training stages. Top row shows average reward curves for (\textbf{top-left})\textbf{ Plan Agent} with expert samples and (\textbf{top-right}) \textbf{Analysis-Answer Agent}. Bottom row presents (\textbf{bottom-left}) \textbf{Rewrite Agent} training dynamics and (\textbf{bottom-right}) policy gradient variance reduction, with the shaded region highlighting the expert injection impact.}
\label{fig:training_curves}
\end{figure}

      \begin{figure}[ht!]
    \centering
    \includegraphics[width=0.48\textwidth]{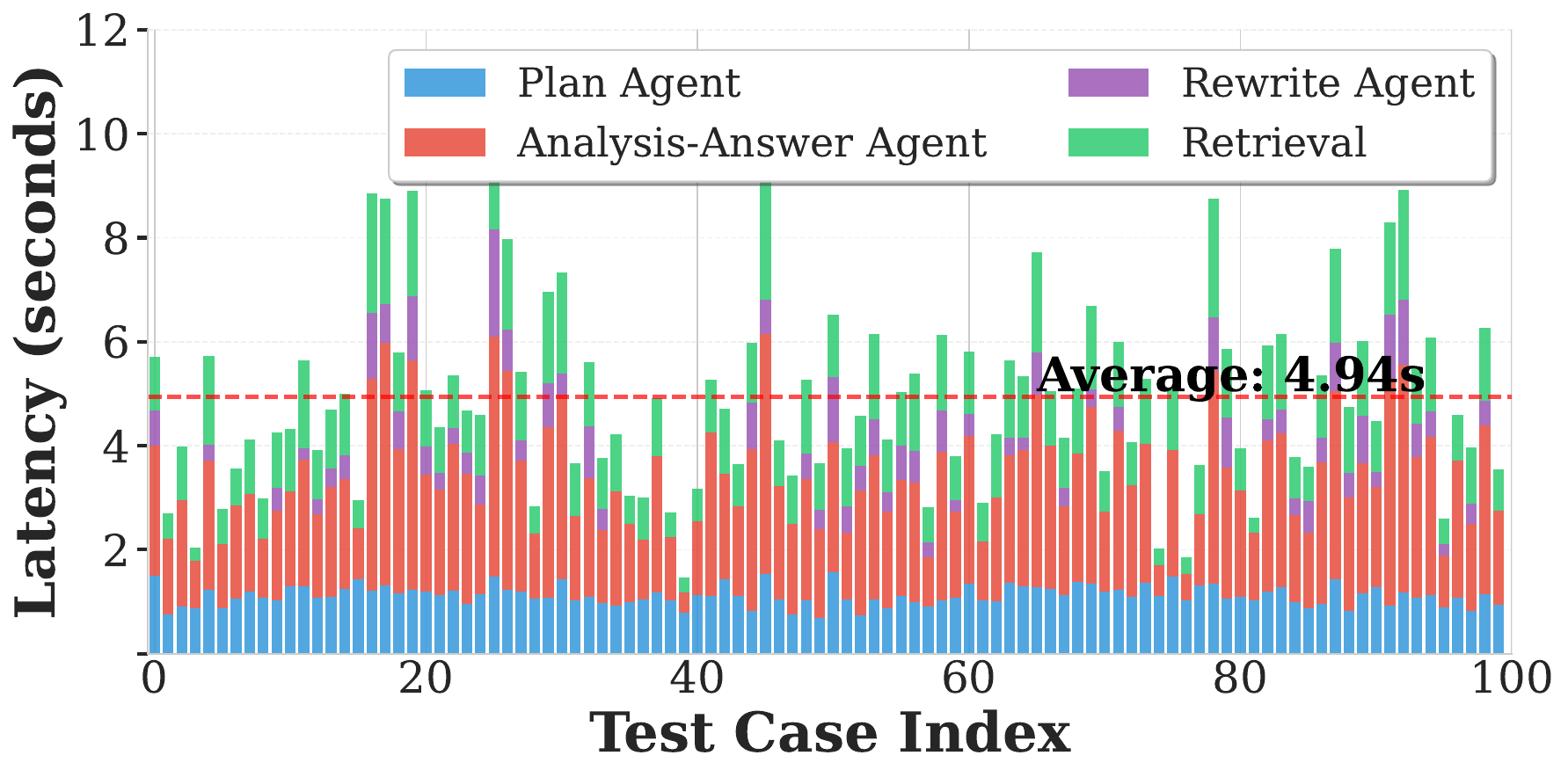}
    \caption{Component-wise latency analysis (100 \textbf{random} questions test)}
    \label{fig:latency_analysis}
    \end{figure}
\noindent\textbf{Consistent Improvements.} OPERA achieves 57.3\% EM on HotpotQA (versus 45.7\% best baseline), 60.2\% EM on 2WikiMultiHopQA (versus 44.3\%), and 39.7\% EM on Musique (versus 24.3\%). These results span different reasoning patterns—comparison, entity traversal, and compositional reasoning—showing the approach works across varied multi-hop tasks.  
     \subsection{Trajectory and Training Analysis}

\noindent \textbf{Runtime Dynamics and Attention Flow.}
Figure~\ref{fig:runtime_dynamics} shows OPERA's decision-making and execution patterns. The top panel shows agent activation over time: the Plan Agent (blue) initiates question cycles, the Analysis-Answer Agent (green) performs reasoning, and the Rewrite Agent (orange) activates upon retrieval failures. The black trajectory tracks the cumulative question completion rate, demonstrating performance across three questions of varying complexity. The bottom panel illustrates attention evolution across processing stages—from entity-focused planning to relation-aware analysis and context-integrated refinement.

\noindent\textbf{Stable Convergence and Reward Evolution.} In reward-driven GRPO, each training step samples eight diverse gold/silver candidates per input, enabling contrastive learning that steers the policy toward better outputs. These signals help agents reach or even surpass $c_{\text{best}}$ (Eq. 11). MAPGRPO enables stable and efficient training with consistent performance improvements. Figure~\ref{fig:training_curves} shows typical RL characteristics: initial instability in the early steps, followed by progressive improvement with occasional dips, particularly visible in the Rewrite Agent due to its conditional activation. Our expert demonstration strategy (Expert Injection Impact 0$\sim$25\%) is highly effective at reducing policy gradient variance, a key factor in training stability. The shaded region in the bottom-right panel highlights this effect, showing a significant variance reduction for the 7B model with expert injection compared to the no-expert variant, confirming our theoretical analysis. The Rewrite Agent is more unstable because it activates only on retrieval failures, yielding sparse rewards; retriever limitations further make Golden-Documents inherently difficult to obtain. Our convergence analysis in Appendix A.5 guarantees only local (not global) convergence and thus no performance guarantee (e.g., Musique EM remains $<$40\% in Table~\ref{tab:main_results}).


    \subsection{Performance Analysis}
     \noindent\textbf{Latency Analysis Across Questions.}
    To analyze latency variations across questions of varying complexity, we evaluate 100 multi-hop test questions, focusing on agent call patterns and component contributions. Figure~\ref{fig:latency_analysis} shows that the Plan Agent maintains relatively consistent latency, while the Analysis-Answer Agent shows higher variance depending on reasoning complexity. The Rewrite Agent activates only when retrieval failures occur, confirming OPERA's adaptive behavior and efficiency for deployment.

\noindent\textbf{Out-of-Domain Evaluation.}
Table~\ref{tab:ood_evaluation} shows how training methods affect generalization. On NQ---single-hop QA over Wikipedia pages where planning is unnecessary---MAPGRPO achieves 36.6\% EM while SFT drops to 19.5\% (from 23.1\% CoT baseline). MAPGRPO preserves OPERA's flexibility, allowing it to bypass planning for single-hop queries and use the Analysis-Answer Agent's training on (sub-question, document, answer) tuples to handle long documents. SFT overfits to multi-hop patterns, attempting decomposition even when not needed. On MHRAG, which has multi-hop structure similar to training data, all methods improve, with MAPGRPO reaching 55.7\% EM. RL methods perform well on both single- and multi-hop tasks, whereas SFT performs worse on single-hop tasks. This suggests that trajectory-based optimization enables adaptive reasoning, while SFT induces rigid behavior.

 \begin{table}[t]
    \centering

    \footnotesize
    \begin{tabular}{lcccc}
    \toprule
    \multirow{2}{*}{\textbf{Method}\textsuperscript{a}} & \multicolumn{2}{c}{\textbf{NQ}} & \multicolumn{2}{c}{\textbf{MHRAG}} \\
    \cmidrule{2-5}
     & EM (\%) & F1 (\%) & EM (\%) & F1 (\%) \\
    \midrule
    CoT  & 23.1 & 29.5 & 42.3 & 51.7 \\
    SFT & 19.5{\tiny{(-3.6)}} & 23.8{\tiny{(-5.7)}} & 44.6{\tiny{(+2.3)}} & 54.2{\tiny{(+2.5)}} \\
    GRPO & 35.6{\tiny{(+12.5)}} & 43.9{\tiny{(+14.4)}} & 50.9{\tiny{(+8.6)}} & 59.5{\tiny{(+7.8)}} \\
    \textbf{MAPGRPO} & \textbf{36.6}{\tiny\textbf{(+13.5)}} & \textbf{45.1}{\tiny\textbf{(+15.6)}} & \textbf{55.7}{\tiny\textbf{(+13.4)}} & \textbf{63.8}{\tiny\textbf{(+12.1)}} \\
    \bottomrule
    \end{tabular}
    \begin{flushleft}
    \footnotesize
    \textsuperscript{a}All methods use the same OPERA architecture. 
    \end{flushleft}
        \caption{Out-of-domain evaluation on single-hop (NaturalQuestions) and multi-hop patterns (MultiHopRAG).}
        \label{tab:ood_evaluation}
\end{table}

\section{Conclusion}
We propose OPERA, a multi-agent framework that addresses limitations in RAG systems through specialized planning and execution roles. OPERA shows improvements—reaching 39.7\% EM on Musique (63.4\% relative improvement) and exceeding 60\% EM on 2WikiMultiHopQA—by combining architectural design and MAPGRPO training, where specialized agents provide separation of concerns while role-specific rewards enable coordination. Ablation studies show that architectural design impacts performance more than training, with Plan Agent removal dropping below untrained baselines. This suggests that reasoning-driven retrieval benefits from architectural advances beyond optimization. While OPERA still struggles with ambiguous decomposition and long reasoning chains, its generalization to out-of-domain tasks demonstrates robustness across reasoning patterns.

\section*{Acknowledgments}
This research is supported by the National Key R\&D Program of China through grant 2023YFC3303800 and Procurement Project through grant E5V01511D3, the National Natural Science
Foundation of China (No. 62406161), the China Postdoctoral Science Foundation (No. 2023M741950) and the Postdoctoral Fellowship Program of CPSF (No. GZB20230347).

\bibliography{aaai2026}

\clearpage
\appendix
\section{Appendix}
    
\subsection{Case Study}
   \noindent \textbf{OPERA vs. Traditional RAG.} We analyze a case study on a query requiring specificity—a common failure point for monolithic RAG systems (Table~\ref{tab:case_study_enhanced_2}). This example shows how OPERA handles ambiguous queries: the Plan Agent sets a clear path, and when initial retrieval proves ambiguous, the Analysis-Answer Agent's failure analysis enables the Rewrite Agent to craft a more specific query.
 This adaptive process resolves the ambiguity. In contrast, the traditional RAG system, lacking specialized roles, fails to bridge the query's distinct concepts, leading to a misinterpretation and an incorrect answer.

    
    
    \begin{table}[!b]
    \centering
    \footnotesize
    \begin{tabular}{p{0.22\columnwidth} p{0.72\columnwidth}}
    \toprule
    \multicolumn{2}{p{0.95\columnwidth}}{\textbf{Question:} Which actor, known for playing a wizard in a famous film series, also starred in the movie "The Good Liar"?} \\
    \midrule
    \multicolumn{2}{l}{\cellcolor{blue!15}\textbf{OPERA: GPM (single agent) with REM (dual-agent)}} \\
    \midrule
    \textbf{GPM}& \textbf{Decomposition:} 1) Which actor played a wizard in a famous film series? \\
    \textbf{Plan Agent}& 2) Did [actor from step 1] star in the movie "The Good Liar"? \\
    \midrule
    \textbf{REM} & \textbf{Step 1 - Initial Fail:} Broad query ``actor who played wizard'' retrieves many actors (e.g., from Harry Potter, Lord of the Rings). \\
    \textbf{Analysis-Answer} & Analysis-Answer Agent: \textbf{NO}, multiple actors match, info is ambiguous. \\
    \cmidrule{2-2}
    \textbf{Agent} & \textbf{Step 1 - Rewrite:} Rewrite Agent receives ambiguity analysis. \\
    \textbf{Rewrite Agent} & New, specific query: ``Ian McKellen Gandalf Lord of the Rings'' OR ``Daniel Radcliffe Harry Potter'' \\
    \cmidrule{2-2}
    \textbf{Analysis-Answer} & \textbf{Step 1 - Success:} Re-retrieval confirms Ian McKellen and Daniel Radcliffe as primary candidates. The system uses the \textit{The Good Liar} cast constraint to resolve the ambiguity. \\
    \textbf{Agent} & Analysis-Answer Agent: \textbf{YES} $\rightarrow$ Ian McKellen \\
    \cmidrule{2-2}
    & \textbf{Step 2:} Analysis-Answer Agent queries ``Did Ian McKellen star in The Good Liar?'' $\rightarrow$ \textbf{YES} \\
    \midrule
    \textbf{Result} & \cellcolor{green!20}\textbf{\checkmark CORRECT:} Ian McKellen. \\
    \midrule
    \multicolumn{2}{l}{\cellcolor{red!15}\textbf{Traditional RAG: Monolithic Processing}} \\
    \midrule
    \textit{Single Query} & ``actor who played wizard in The Good Liar'' \\
    \midrule
    \textit{Processing} & Retrieves docs for "The Good Liar" and general pages about "wizards". \\
    & \textbf{Critical Flaw:} Fails to bridge the two concepts ("wizard actor" + "The Good Liar cast"). The model misinterprets the query's core intent. \\
    \midrule
    \textbf{Result} & \cellcolor{red!20}\textbf{$\times$ INCORRECT:} The main actor in "The Good Liar" is Helen Mirren. The query about a wizard is likely incorrect. \\
    \bottomrule
    \end{tabular}
        \caption{Case Study: OPERA vs. Traditional RAG on a Query Requiring Specificity}
    \label{tab:case_study_enhanced_2}
    \end{table}

    \begin{table}[H]
    \centering
    \scriptsize
    \begin{tabular}{p{0.18\columnwidth} p{0.76\columnwidth}}
    \toprule
    \multicolumn{2}{p{0.95\columnwidth}}{\textbf{Question:} What is the population of the city where the company that acquired WhatsApp is headquartered?} \\
    \midrule
    \multicolumn{2}{l}{\cellcolor{blue!15}\textbf{Step 1: PLAN Agent (7B) - Strategic Decomposition}} \\
    \midrule
    \textbf{Generated Plan} & 1) Which company acquired WhatsApp? \\
    & 2) Where is [company from step 1] headquartered? \{deps: [1]\} \\
    & 3) What is the population of [city from step 2]? \{deps: [2]\} \\
    & \textcolor{green!60!black}{Reward: 0.92 (High-quality decomposition)} \\
    \midrule
    \multicolumn{2}{l}{\cellcolor{orange!15}\textbf{Step 2: Analysis-Answer Agent - Subgoal 1}} \\
    \midrule
    \textbf{Query} & ``Which company acquired WhatsApp?'' \\
    \textbf{Retrieved} & 1. ``...Facebook Inc. announced the acquisition of WhatsApp for \$19 billion...'' \\
    & 2. ``...The messaging app WhatsApp, owned by Meta (formerly Facebook)...'' \\
    \textbf{Analysis} & Clear evidence that Facebook (now Meta) acquired WhatsApp. \\
    \textbf{Decision} & \textcolor{green!60!black}{\textbf{YES}} $\rightarrow$ Answer: ``Facebook (now Meta)'' (Conf: 0.95) \\
    \textbf{TMC Update} & \textcolor{blue!60!black}{Entity Resolution: ``company from step 1'' $\rightarrow$ ``Facebook/Meta''} \\
    \midrule
    \multicolumn{2}{l}{\cellcolor{orange!15}\textbf{Step 3: Analysis-Answer Agent - Subgoal 2}} \\
    \midrule
    \textbf{Query} & ``Where is Facebook/Meta headquartered?'' \\
    \textbf{Retrieved} & 1. ``...Meta Platforms, Inc. is headquartered in Menlo Park, California...'' \\
    & 2. ``...Facebook's main campus is located at 1 Hacker Way, Menlo Park, CA...'' \\
    \textbf{Decision} & \textcolor{green!60!black}{\textbf{YES}} $\rightarrow$ Answer: ``Menlo Park, California'' (Conf: 0.98) \\
    \textbf{TMC Update} & \textcolor{blue!60!black}{Entity Resolution: ``city from step 2'' $\rightarrow$ ``Menlo Park''} \\
    \midrule
    \multicolumn{2}{l}{\cellcolor{red!10}\textbf{Step 4: Analysis-Answer Agent - Subgoal 3 (Initial Failure)}} \\
    \midrule
    \textbf{Query} & ``What is the population of Menlo Park?'' \\
    \textbf{Retrieved} & 1. ``...Menlo Park is known for its tech companies...'' \\
    & 2. ``...The city has numerous parks and recreational areas...'' \\
    \textbf{Analysis} & Retrieved documents lack specific population data. \\
    \textbf{Decision} & \textcolor{red!60!black}{\textbf{NO}} - Need more specific information \\
    \midrule
    \multicolumn{2}{l}{\cellcolor{purple!15}\textbf{Step 5: Rewrite Agent (3B) - Query Refinement}} \\
    \midrule
    \textbf{Failed Query} & ``What is the population of Menlo Park?'' \\
    \textbf{Strategy} & Add demographic keywords and year specificity \\
    \textbf{Rewritten} & ``Menlo Park California population census demographics statistics 2020 2023'' \\
    \textbf{Keywords} & [``population'', ``census'', ``demographics'', ``statistics'', ``residents''] \\
    \midrule
    \multicolumn{2}{l}{\cellcolor{green!10}\textbf{Step 6: Analysis-Answer Agent - Subgoal 3 (Success)}} \\
    \midrule
    \textbf{Query} & ``Menlo Park California population census demographics statistics 2020 2023'' \\
    \textbf{Retrieved} & 1. ``...According to the 2020 census, Menlo Park had a population of 35,211...'' \\
    & 2. ``...Menlo Park demographics show a diverse community with 35,211 residents...'' \\
    \textbf{Decision} & \textcolor{green!60!black}{\textbf{YES}} $\rightarrow$ Answer: ``35,211 (as of 2020 census)'' (Conf: 0.93) \\
    \midrule
    \textbf{Final Answer} & \cellcolor{green!20}The population of Menlo Park, California, where Meta (formerly Facebook), the company that acquired WhatsApp, is headquartered, is 35,211 according to the 2020 census. \\
    \midrule
    \textbf{Metrics} & Steps: 6 | Plan: 1 | Analysis: 4 | Rewrite: 1 | Success: 75\% | Latency: 5.7s \\
    \bottomrule
    \end{tabular}
    \caption{Complete OPERA trajectory showing collaborative multi-agent reasoning with TMC coordination.}
    \label{tab:trajectory_case}
    \end{table}
      \noindent \textbf{ Complete Trajectory.} To illustrate OPERA's complete reasoning process, we present a detailed trajectory for a complex multi-hop question from the Musique dataset. This case demonstrates how our three agents collaborate through the Trajectory Memory Component (TMC) to solve a challenging query.
    \subsection{Training Process Settings}
    
\begin{figure}[t]
    \centering
    \includegraphics[width=0.48\textwidth]{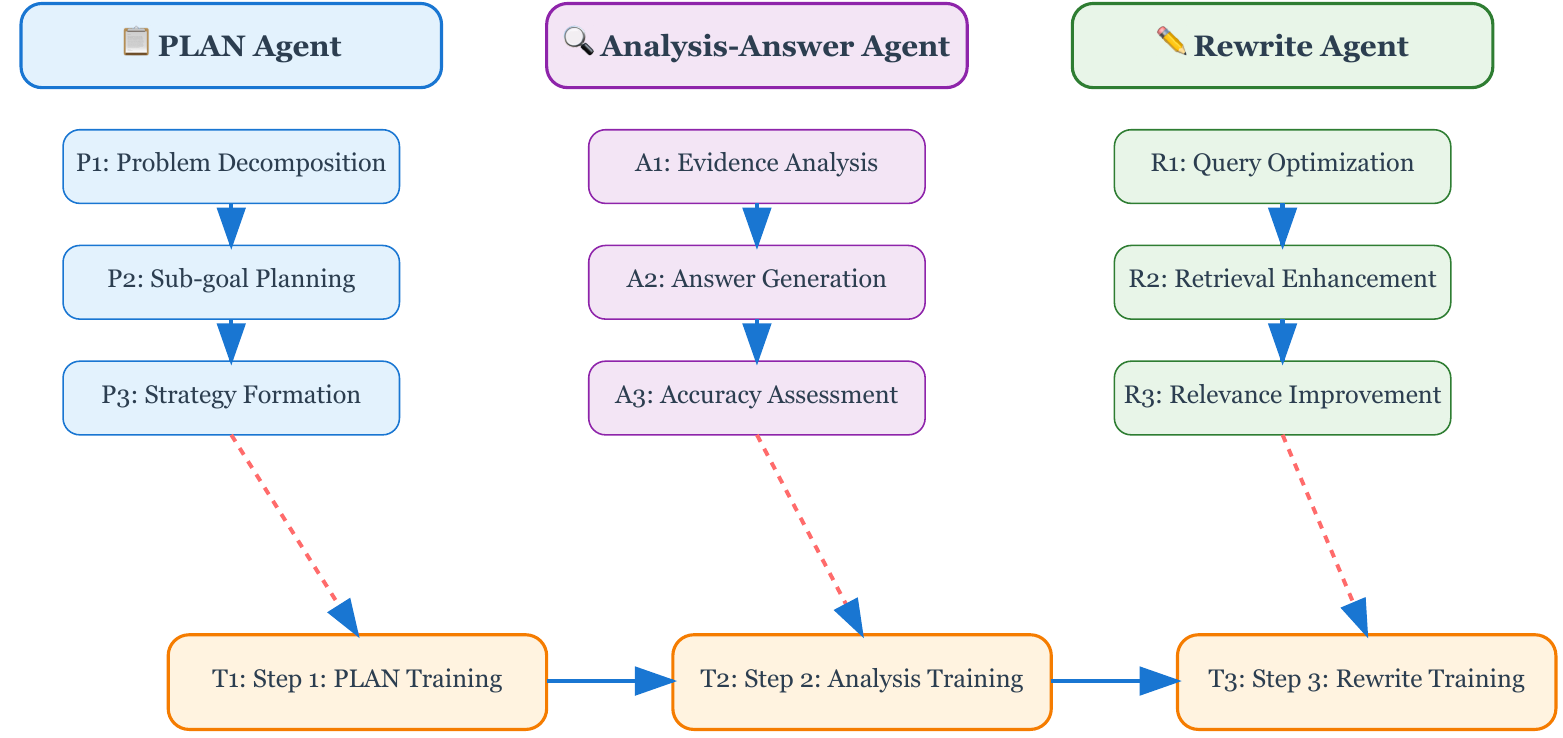}
    \caption{MAPGRPO training pipeline illustrating the three-stage sequential optimization process. Each stage focuses on a specific agent with tailored reward functions.}
    \label{fig:mapgrpo_pipeline}
    \end{figure}

    \noindent \textbf{MAPGRPO Training Pipeline.}
    Figure \ref{fig:mapgrpo_pipeline} illustrates the Multi-Agents Progressive Group Relative Policy Optimization (MAPGRPO) training pipeline, a three-stage sequential process designed to specialize our core agents. This modular approach applies tailored reward functions at each stage, with each agent specializing in its distinct role. The pipeline begins with Stage 1, where the PLAN Agent learns problem decomposition, sub-goal planning, and strategy formation. Its outputs then train the Analysis-Answer Agent in Stage 2, which learns to execute plans through evidence analysis, answer generation, and accuracy assessment. Finally, in Stage 3, the Rewrite Agent learns to optimize queries, improve retrieval, and increase relevance for adaptive error correction. This progressive specialization enables OPERA's reasoning capabilities. In the Plan Agent stage, the red-star markers in Figure~\ref{fig:training_curves} denote reference-refresh points for $c_{\text{best}}$; the transient reward drops after these points reflect a harder group-level comparison baseline before the policy adapts. This refresh mechanism prevents the policy from overfitting to weak within-group comparisons by periodically reintroducing a stronger reference anchor.
    
    \noindent \textbf{Training Data Format and Paradigm.}
    Different methods use different training approaches. The training paradigms are:
    \begin{itemize}
        \item \textbf{External Baseline Methods (Adaptive-RAG, BGM):} These methods are trained end-to-end on complete multi-hop questions and their final answers, learning to directly map complex questions to ultimate answers.
        \item \textbf{OPERA Variants (CoT, SFT, GRPO, MAPGRPO):} All OPERA configurations, including ablation variants, are trained on decomposed sub-problems. Specifically:
        \begin{itemize}
            \item The Plan Agent is trained to decompose multi-hop questions into sequential sub-goals
            \item The Analysis-Answer Agent is trained on (sub-question, retrieved documents, sub-answer) tuples, where sub-questions are atomic queries answerable from a small document set
            \item The Rewrite Agent is trained to reformulate failed sub-queries for better retrieval
        \end{itemize}
    \end{itemize}
    The consistent use of decomposed training across all OPERA variants (including the SFT and GRPO ablations) enables fair comparison and shows that performance gains come from the training methodology rather than the data format. In the GRPO training groups, we refer to online rollout candidates generated by the current policy as silver candidates, while the injected offline high-score reference candidate $c_{\text{best}}$ is treated as the gold candidate. In Figure~\ref{fig:motivation}, ``judge'' denotes the reward-scoring source for each agent, including online rollout scoring, offline labels, rule-based checks, and execution-based signals.
    
    \noindent \textbf{Training Hyperparameters.}
    Our MAPGRPO training uses different hyperparameters for each agent, tailored to their specific roles in the OPERA framework. As shown in Table~\ref{tab:hyperparameters}, we use different learning rates and reward component weights ($\lambda_i, \alpha, \beta, \gamma, \omega_i$) to match the objectives of the Plan, Analysis-Answer, and Rewrite agents. For instance, the Analysis-Answer agent's reward emphasizes exact match accuracy ($\beta=0.65$), while the Rewrite agent focuses on retrieval effectiveness ($\omega_1=0.9$). This configuration supports optimal performance in our progressive training pipeline.

\begin{table}[t]
    \centering
    \footnotesize
    \begin{tabular}{lccc}
    \toprule
    \textbf{Agent Parameter} & \textbf{Plan} & \textbf{Analysis-Answer} & \textbf{Rewrite} \\
    \midrule
    lr & 5e-6 & 1e-5 & 2e-5 \\
    bs & 8 & 8 & 8 \\
    grad\_accum & 4 & 4 & 4 \\
    eff\_bs & 32 & 32 & 32 \\
    epochs & 3 & 2 & 2 \\
    train\_samples & 35K & 50K & 35K \\
    $\lambda_1$ / $\alpha$ & 0.4 & 0.25 & - \\
    $\lambda_2$ / $\beta$ & 0.3 & 0.65 & - \\
    $\lambda_3$ / $\gamma$ & 0.3 & 0.1 & - \\
    kl\_coef ($\beta$) & 0.1 & 0.1 & 0.1 \\
    group\_sz ($G$) & 8 & 8 & 8 \\
    $\omega_1$ / $\omega_2$ & - & - & 0.9 / 0.1 \\
    \bottomrule
    \end{tabular}
    \caption{MAPGRPO training hyperparameters for each agent.}
    \label{tab:hyperparameters}
\end{table}
    
    \noindent \textbf{Baseline Method Hyperparameters.}
    We configured baseline methods using their officially published hyperparameters or conducted grid search to optimize their performance on our benchmarks. Table~\ref{tab:baseline_hyperparams} lists the key settings for Adaptive-RAG  and BGM . These configurations provide competitive implementations of each baseline.

  \begin{table}[t]
  \centering
  \footnotesize
  \begin{tabular}{lcc}
  \toprule
  \textbf{Method} & \textbf{Parameter} & \textbf{Value} \\
  \midrule
  \multirow{5}{*}{Adaptive-RAG} & Classifier LR & 5e-4 \\
   & LoRA Rank & 16 \\
   & LoRA Alpha & 32 \\
   & Target Modules & q\_proj, v\_proj \\
   & Batch Size & 32 \\
  \midrule
  \multirow{5}{*}{BGM} & Bridge LR & 5e-6 \\
   & PPO \cite{schulman2017proximal} LR & 1e-5 \\
   & Batch Size & 16 \\
   & PPO Epochs & 3 \\
   & Entropy Coeff. & 0.0 \\
  \bottomrule
  \end{tabular}
  \caption{Hyperparameters for baseline methods.}
  \label{tab:baseline_hyperparams}
  \end{table}

\begin{table*}[t!]
\centering
\footnotesize
\renewcommand{\arraystretch}{0.9}
\begin{tabular}{l|cc|cc|cc}
\toprule
\multirow{2}{*}{\textbf{Method}} & \multicolumn{2}{c|}{\textbf{HotpotQA}} & \multicolumn{2}{c|}{\textbf{2WikiMultiHopQA}} & \multicolumn{2}{c}{\textbf{Musique}} \\
\cmidrule{2-7}
 & EM (\%) & F1 (\%) & EM (\%) & F1 (\%) & EM (\%) & F1 (\%) \\
\midrule
Qwen2.5-7B-Instruct\textsuperscript{a} + OPERA CoT & 44.9 & 58.5 & 42.3 & 50.7 & 21.2 & 32.1 \\
LLama3.1-8B-Instruct\textsuperscript{b} + OPERA CoT & 38.7 & 48.3 & 35.6 & 45.1 & 18.3 & 23.2 \\
Qwen3-8B\textsuperscript{c} + OPERA CoT & 47.7 & 61.4 & 49.3 & 62.8 & 30.2 & 39.1 \\
LLama-3.1-70B-Instruct\textsuperscript{b} + OPERA CoT & 54.9 & 65.8 & 52.8 & 64.7 & 34.1 & 45.2 \\
\midrule
\textbf{OPERA (MAPGRPO)} & \textbf{57.3} & \textbf{69.5} & \textbf{60.2} & \textbf{72.7} & \textbf{39.7} & \textbf{58.0} \\
\bottomrule
\end{tabular}

\noindent\footnotesize \textsuperscript{a}Qwen2.5 series~\cite{yang2024qwen2.5}. \textsuperscript{b}LLama3.1 series~\cite{grattafiori2024llama3}. \textsuperscript{c}Qwen3~\cite{yang2025qwen3}.
\caption{Performance comparison of OPERA across different open-source model scales.}
\label{tab:model_scale_comparison}
\end{table*}

\label{subsec:corpus_settings}
\noindent \textbf{Corpus Settings.} We construct a unified retrieval corpus solely from the paragraph
collections released with HotpotQA~\cite{yang2018hotpotqa},
2WikiMultiHopQA~\cite{ho2020constructing}, and
MusiQue~\cite{trivedi2022musique}, without external Wikipedia pages.
The corpus is centered on MusiQue while jointly accommodating the other
two datasets: MusiQue's gold supporting evidence is embedded within a
denser, in-distribution distractor pool drawn from HotpotQA and
2WikiMultiHopQA, yielding a more challenging open-domain setting. We
apply sentence-level indexing with exact $(\text{title},\text{content})$
deduplication, which preserves gold evidence without loss while removing
redundant chunks across datasets. Each unique sentence is embedded with
BGE-M3~\cite{chen2024bge} and indexed with an exact FAISS inner-product
(cosine) index~\cite{johnson2019billion}. As shown in
Table~\ref{tab:corpus_stats}, deduplication reduces 2.09M raw chunks to
\textbf{1.78M} unique chunks (about 15\% removed, predominantly
cross-dataset duplicates). During evaluation on any single dataset,
retrieval runs over the entire unified corpus, so relevant evidence must
be surfaced from distractors contributed by all three sources.

\begin{table}[t]
\centering
\begin{tabular}{lr}
\toprule
\textbf{Source} & \textbf{Unique Chunks} \\
\midrule
HotpotQA & 269,558 \\
2WikiMultiHopQA & 1,758,475 \\
MusiQue & 57,112 \\
\midrule
\textbf{Sum Before Cross-Dataset Dedup} & \textbf{2.09M (2,085,145)} \\
\textbf{Final After Cross-Dataset Dedup} & \textbf{1.78M (1,780,294)} \\
\bottomrule
\end{tabular}
\caption{Statistics of the unified retrieval corpus showing cross-dataset deduplication effects. Per-source counts aggregate the paragraph context available in our ingestion setup.}
\label{tab:corpus_stats}
\end{table}

\subsection{Additional Results}

\noindent \textbf{OPERA Architecture Evaluation Across Model Scales.} Table \ref{tab:model_scale_comparison} evaluates OPERA's performance across different open-source models. The architecture shows benefits across model scales: Qwen2.5-7B achieves 44.9\% EM on HotpotQA with OPERA CoT, while LLama-3.1-70B reaches 54.9\% EM. Among the 8B-scale models tested, Qwen3-8B with OPERA CoT achieves 47.7\% EM on HotpotQA and 30.2\% EM on Musique, showing that the architecture can work with models of different sizes and families.

\noindent \textbf{MAPGRPO Training Impact.} Comparing OPERA CoT and full MAPGRPO under the same Qwen2.5-7B backbone reveals consistent 
gains from training: 12.4\% EM on HotpotQA (44.9 $\to$ 57.3), 17.9\% EM on 2WikiMultiHopQA 
(42.3 $\to$ 60.2), and 18.5\% EM on Musique (21.2 $\to$ 39.7). Notably, the 7B model trained 
with MAPGRPO surpasses LLama-3.1-70B with OPERA CoT across all benchmarks, suggesting that 
specialized training can outweigh raw model scale. The larger improvements on 2WikiMultiHopQA and Musique suggest that the specialized training may be more beneficial for complex multi-hop reasoning tasks. These results indicate that both architectural design and progressive training contribute to the overall performance gains.

        \noindent\textbf{Rewrite Agent Model Scale Analysis.}
We evaluated different model sizes for the Rewrite Agent to find the optimal balance between performance and efficiency. Table~\ref{tab:scale_ablation} shows that while the 7B model provides a marginal 0.6\% EM improvement over the 3B model, it introduces 0.6s additional latency per question. Conversely, the 1.5B model, though slightly faster, suffers a substantial drop in performance. Therefore, the Qwen2.5-3B model delivers robust performance with zero latency overhead, making it the clear choice for our architecture.

    \begin{table}[t]
    \centering
    \small
    \begin{tabular}{lcccc}
    \toprule
    \textbf{Rewrite-Agent Model} & \textbf{EM} & \textbf{F1} & \textbf{nDCG} & \textbf{Extra} \\
     & \textbf{(\%)} & \textbf{(\%)} & \textbf{@10} & \textbf{(s)} \\
    \midrule
    Qwen2.5-7B-Instruct & 57.9 & 70.0 & 0.82 & 0.6 \\
    \textbf{Qwen2.5-3B-Instruct} & \textbf{57.3} & \textbf{69.5} & \textbf{0.79} & \textbf{0.0} \\
    Qwen2.5-1.5B-Instruct & 50.8 & 63.7 & 0.68 & -0.3 \\
    \bottomrule
    \end{tabular}
        \caption{Impact of model scale choices for the Rewrite Agent on end-to-end performance.}
    \label{tab:scale_ablation}
    \end{table}

\noindent\textbf{Document Scaling Analysis.} Table~\ref{tab:document_scaling} reveals that K=5 provides the optimal balance between performance and efficiency. While increasing K to 10 yields marginal improvements (0.8\% EM), it significantly increases noise (measured as the ratio of irrelevant documents), justifying our choice of Top-5 retrieval. Note that these experiments use only the base retrieval and Analysis-Answer components without the full OPERA pipeline.
\begin{table}[t]
\centering
\begin{tabular}{lccc}
\toprule
\textbf{Top-K} & \textbf{EM (\%)} & \textbf{F1 (\%)} & \textbf{Noise Score} \\
\midrule
3 & 46.8 & 58.9 & 0.15 \\
\textbf{5} & \textbf{50.4} & \textbf{63.0} & \textbf{0.22} \\
10 & 51.2 & 63.8 & 0.35 \\
15 & 51.0 & 63.5 & 0.48 \\
20 & 50.2 & 62.9 & 0.62 \\
\bottomrule
\end{tabular}
\caption{Impact of retrieved document count (Top-K) on performance and efficiency.}
\label{tab:document_scaling}
\end{table}
    
    \noindent \textbf{Error Analysis.}
To understand OPERA's failure modes, we analyzed error distributions over 10 random samples of 200 failed cases from each dataset and report the averaged percentages. Table~\ref{tab:error_analysis} presents a detailed breakdown of error types, where primary categories are mutually exclusive (one per case). Detailed labels are annotated independently and may co-occur or overlap across categories. The analysis reveals dataset-specific patterns: HotpotQA and 2WikiMultiHopQA failures are predominantly due to reasoning errors (62.1\% and 69.2\%), particularly incorrect YES decisions where the Analysis-Answer Agent mistakenly believes it has sufficient information. In contrast, Musique shows a more balanced distribution with retrieval errors accounting for 47.8\% of failures, reflecting its more complex multi-hop nature that challenges our retrieval system even with the Rewrite Agent's assistance.

    \begin{table}[t]
    \centering
    \small
    \begin{tabular}{lccc}
    \toprule
    \textbf{Error Type} & \textbf{HotpotQA} & \textbf{2Wiki} & \textbf{Musique} \\
    \midrule
    \multicolumn{4}{l}{\textit{Primary Error Sources (\% of failed cases)}} \\
    \midrule
    Planning Errors & 13.3 & 11.1 & 14.1 \\
    Retrieval Errors & 24.6 & 19.7 & 47.8 \\
    Reasoning Errors & 62.1 & 69.2 & 38.1 \\
    \midrule
    \multicolumn{4}{l}{\textit{Detailed Breakdown (\% of failed cases)}} \\
    \midrule
    Incorrect decomposition & 8.2 & 6.8 & 9.5 \\
    Missing dependencies & 5.1 & 4.3 & 4.6 \\
    Retrieval despite rewrite & 21.7 & 17.3 & 32.5 \\
    Retrieval without rewrite & 7.3 & 6.8 & 10.4 \\
    Incorrect YES decision & 42.1 & 48.3 & 31.7 \\
    Misextraction from context & 16.8 & 17.5 & 9.7 \\
    \bottomrule
    \end{tabular}
        \caption{Error distribution analysis across agent behaviors.}
    \label{tab:error_analysis}
    \end{table}
    
\begin{figure}[t]
\centering
\includegraphics[width=0.48\textwidth]{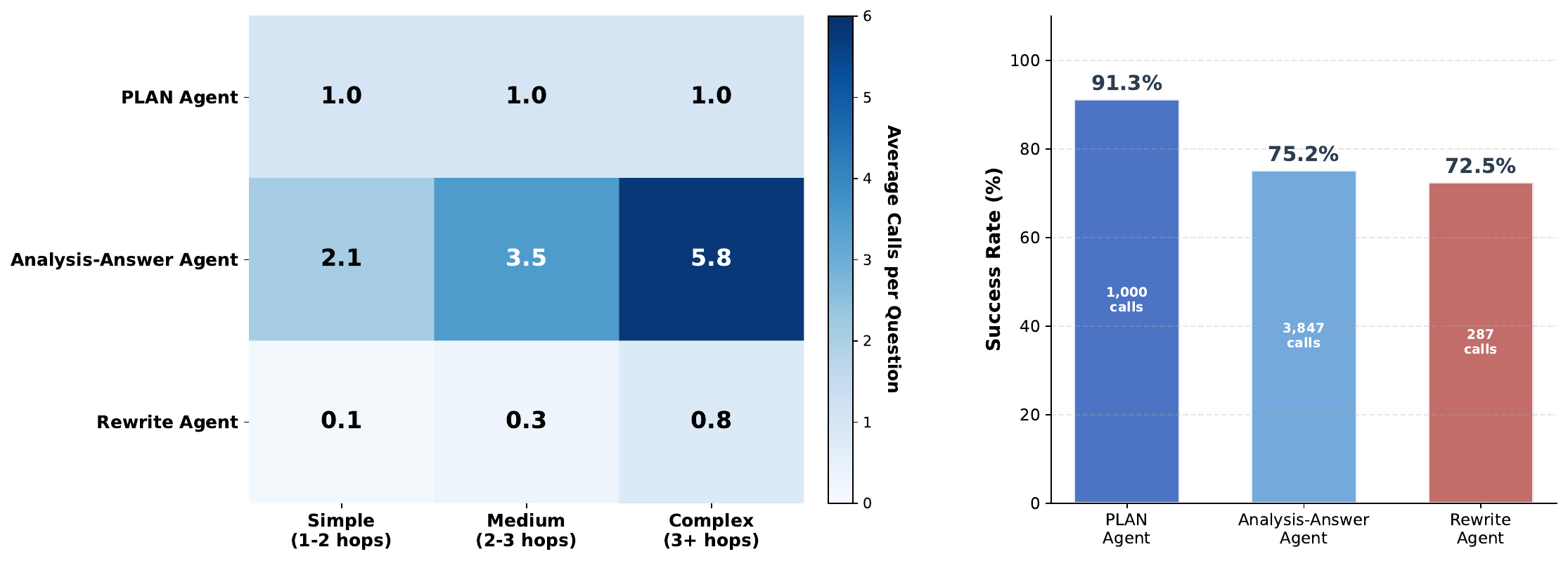}
\caption{(Left) Heatmap of average agent calls per question, categorized by complexity. (Right) Execution success rates for each agent across 1,000 test questions.}
\label{fig:call_patterns}
\end{figure}

\noindent\textbf{Call Patterns vs. Complexity.} 
OPERA dynamically allocates resources based on question difficulty. As shown in Figure~\ref{fig:call_patterns} (left), average agent calls increase with complexity; for instance, the Analysis-Answer agent's calls rise from 2.1 on simple questions to 5.8 on complex ones, while the Rewrite agent's calls climb from 0.1 to 0.8. This confirms OPERA's adaptive reasoning process. Figure~\ref{fig:call_patterns} (right) complement this by showing high execution success rates (91.3\% for Plan, 72.5\% for Rewrite), affirming the reliability of each component and quantifying their contribution to overall performance.

\noindent\textbf{Success and Pattern Analysis.} To complement the error analysis, we analyzed 10 random samples of 100 successful cases from each dataset and report the averaged percentages in Table~\ref{tab:success_patterns}. The data reveals that simpler datasets (HotpotQA, 2Wiki) achieve higher success rates through direct retrieval, while complex multi-hop questions (Musique) more frequently require the Rewrite Agent. Rewrite efficiency measures the percentage of rewrites that led to finding the correct answer.

    \begin{table}[t!]
    \centering
    \small
    \begin{tabular}{lccc}
    \toprule
    \textbf{Success Pattern} & \textbf{HotpotQA} & \textbf{2Wiki} & \textbf{Musique} \\
    \midrule
    Direct retrieval (no rewrite) & 62.3\% & 65.8\% & 47.1\% \\
    Retrieval after rewrite & 26.5\% & 23.7\% & 34.2\% \\
    Multiple rewrites needed & 11.2\% & 10.5\% & 18.7\% \\
    \midrule
    \textbf{Avg hops completed} & 2.1 & 2.3 & 3.4 \\
    \textbf{Avg Analysis calls} & 2.8 & 3.1 & 4.6 \\
    \textbf{Rewrite efficiency} & 67.5\% & 71.2\% & 58.9\% \\
    \bottomrule
    \end{tabular}
        \caption{Success patterns in correctly answered questions.}
    \label{tab:success_patterns}
    \end{table}
    


\subsection{Design Philosophy and Technical Trade-offs}
During the development of OPERA, we encountered several fundamental questions that shaped our design decisions. We present these as a series of questions and answers to illuminate the philosophical and technical considerations underlying our approach.

\noindent\textbf{Q1: Why does the Analysis-Answer Agent's reward function not include explicit metrics for analysis quality?} Our experiments show a strong correlation between answer correctness and reasoning quality in multi-hop questions. Rather than designing complex multi-objective rewards, we optimize solely for answer exactness (EM score). In multi-hop reasoning, producing correct answers requires valid intermediate reasoning steps. This simplified reward structure reduces training complexity while achieving comparable reasoning quality, as our experimental results show.

\noindent\textbf{Q2: Why prioritize prompt engineering over supervised fine-tuning (SFT) for the base agents?}
Prompt engineering provides immediate deployment with minimal computational overhead, requiring only inference-time modifications rather than expensive gradient updates. This approach preserves model generality while SFT creates rigid behaviors that attempt decomposition on simple queries needing only direct retrieval. Prompts achieve 21.2\% EM baseline performance with minimal implementation cost. The simple implementation allows rapid iteration and ablation studies that would be difficult with fine-tuning cycles. By using prompts for behavioral guidance and MAPGRPO for performance optimization, we maintain model flexibility while avoiding the computational cost and overfitting risks of supervised training.

\noindent\textbf{Q3: How does MAPGRPO training differ from traditional fine-tuning in preserving model capabilities?}
MAPGRPO uses sequential agent training where each subsequent agent adapts to outputs from previously trained agents. This ordered training (Plan → Analysis-Answer → Rewrite) ensures downstream agents learn from realistic input distributions rather than idealized data. The approach maintains 36.6\% EM on out-of-domain NaturalQuestions versus 19.5\% for SFT. This performance difference occurs through three mechanisms: \textbf{1)} KL-constrained optimization prevents distribution collapse; \textbf{2}) Group-relative ranking rewards improvement without forcing convergence to single solutions; \textbf{3)} Sequential training allows natural adaptation to predecessor agents' actual behaviors rather than oracle outputs.

    \noindent\textbf{Q4: Why separate planning, analysis, and rewriting into distinct agents rather than training a single unified model?}
Modular decomposition enables independent optimization of specialized capabilities before system integration. Ablation studies show severe performance drops without this separation (39.7\% → 17.1\% EM). The architecture addresses performance bottlenecks that occur in monolithic training where the weakest capabilities limit overall performance. By training each agent to excel in its specific domain—planning on decomposition tasks, analysis on document extraction, and rewriting on query reformulation—then sequentially adapting them, we avoid the severe degradation observed in the w/o Plan Agent ablation, which drops performance by 22.6 EM points. This approach converts a complex multi-objective problem into manageable specialized optimizations.
    
    \noindent\textbf{Q5: What are the implications of relying on answer correctness as a proxy for reasoning quality?}
  This design has both benefits and limitations. The correlation between answer correctness and reasoning validity works well for multi-hop questions requiring sequential inference, since incorrect intermediate steps typically lead to wrong answers. However, this approach may miss cases where correct answers result from spurious correlations rather than valid reasoning chains. Our validation shows the approach adequately captures reasoning quality for practical applications, achieving 39.7\% EM with consistent reasoning paths. The simplified reward signal speeds training convergence while avoiding complex multi-metric optimization that can produce conflicting gradients.
  
     \noindent\textbf{Q6: Why does the ablation ``w/o PLAN Agent'' perform worse than the CoT baseline, despite both using the same underlying models?}
 This unexpected result reveals a training-inference distribution mismatch. The Analysis-Answer Agent trains exclusively on atomic sub-questions with localized document sets. Without the Plan Agent, it faces compound multi-hop queries that it never encountered during training. Performance drops significantly (17.1\% EM) as the agent attempts direct retrieval for complex queries like "What is the birthplace of the director of Inception?" rather than working with decomposed atomic queries. CoT maintains consistency between training and inference distributions through uniform prompt-based decomposition. This result shows that modular architectures require either comprehensive training coverage or architectural guarantees of appropriate input distributions.
      
  \noindent\textbf{Q7: What is the fundamental principle underlying OPERA's design, and why does it enable smaller models to compete with much larger systems?}
  OPERA uses specialized decomposition where complex reasoning emerges from coordinated simple operations rather than monolithic computation. The architecture builds on two observations: First, individual reasoning steps such as "Who directed Inception?" are manageable for smaller models. Second, reasoning difficulty comes from orchestration rather than execution. By separating planning from execution from error recovery, we reduce each component's complexity to match smaller models' capabilities. Using 7B and 3B models, we achieve 39.7\% EM on Musique, showing that architectural design and specialized training can significantly narrow the performance gap with larger systems. This approach indicates a shift toward horizontal scaling through specialization rather than vertical scaling through parameter growth.

\subsection{Detailed Theoretical Analysis}

We analyze OPERA's architecture and training methodology, examining convergence behavior, reward function design, and architectural trade-offs.

\noindent \textbf{Convergence Analysis of MAPGRPO.} MAPGRPO converges to local optima under standard regularity conditions.

\noindent \textbf{Regularity Conditions.} For each agent $k \in \{1,2,3\}$:
\begin{enumerate}
    \item The reward function $r^{(k)}$ is bounded: $|r^{(k)}(x,y)| \leq R_{max}$ for all $(x,y)$.
    \item The policy $\pi_{\theta_k}^{(k)}$ is differentiable with respect to $\theta_k$ and satisfies the Lipschitz condition: $\|\nabla_{\theta_k} \log \pi_{\theta_k}^{(k)}(y|x)\| \leq L$ for some constant $L > 0$.
    \item The KL divergence constraint is satisfied: $E_{x \sim D_k}[D_{KL}[\pi_{\theta_k}^{(k)}(\cdot|x) \| \pi_{ref}^{(k)}(\cdot|x)]] \leq \epsilon_{KL}$ for some $\epsilon_{KL} > 0$.
\end{enumerate}

\noindent \textbf{MAPGRPO Convergence.} Under these conditions, each stage of MAPGRPO converges to a local optimum of its objective function. For agent $k$ trained in stage $k$, the expected squared gradient norm satisfies:
\begin{align}
E\left[\|\nabla_{\theta_k} J_k(\theta_k|\theta^*_{<k})\|^2\right] = O\left(\frac{1}{\sqrt{T_k}}\right),
\end{align}
where $T_k$ is the number of training steps for agent $k$.

\noindent\textbf{Proof Sketch:} We establish convergence for a general agent $k$ through three key steps. 

First, we bound the variance of the group-relative advantage function $A_i^{(k)}(x, y_i) = r^{(k)}(x, y_i) - \frac{1}{G}\sum_{j=1}^G r^{(k)}(x, y_j)$. Since rewards are bounded by $R_{max}$, we have $|A_i^{(k)}(x, y_i)| \leq 2R_{max}$, which directly yields the variance bound $Var[A_i^{(k)}(x, y_i)] \leq 4R_{max}^2$.

Second, we construct the policy gradient estimator as $\hat{g}_k = \frac{1}{B}\sum_{b=1}^B \sum_{i=1}^G A_i^{(k)}(x_b, y_{b,i}) \nabla_{\theta_k} \log \pi_{\theta_k}^{(k)}(y_{b,i}|x_b)$, which incorporates both the bounded advantage function and the Lipschitz gradient condition.

Finally, applying standard policy gradient convergence analysis with our bounded variance and Lipschitz assumptions yields the convergence rate $E\left[\|\nabla_{\theta_k} J_k(\theta_k|\theta^*_{<k})\|^2\right] \leq \frac{C}{\sqrt{T_k}}$, where the constant $C = 8R_{max}^2 L^2 (1 + \epsilon_{KL})$ depends on our regularity conditions.

\noindent \textbf{Reward Function Design.} Our reward functions are motivated by an information-theoretic view of multi-hop reasoning. By the chain rule of mutual information,
\begin{align}
I(Q; (P,A)\mid D) = I(Q; P\mid D) + I(Q; A\mid D, P),
\end{align}
the information a system obtains about the answer factorizes into a planning term $I(Q;P\mid D)$ and a plan-conditioned answering term $I(Q;A\mid D,P)$, while query rewriting contributes an additional gain through the refined document set $D'$. Accordingly, the Plan-Agent reward ($f_{\text{logic}}, f_{\text{struct}}, f_{\text{exec}}$) targets the planning term, the Analysis-Answer reward (sufficiency indicator and exact match) targets the answering term, and the Rewrite-Agent NDCG reward targets the gain attributable to $D'$. We do not claim these rewards are information-theoretically optimal: the component weights ($\lambda_i, \alpha, \beta, \gamma, \omega_i$) are selected by validation grid search (Table~\ref{tab:hyperparameters}), not derived in closed form.

\noindent \textbf{Agent Decomposition Analysis.} A multi-hop reasoning task has complexity $C(Q) = (h, s, r)$ where $h$ is the number of reasoning hops, $s$ is the search space size, and $r$ is the reasoning complexity within each hop. We compare orchestration cost under a coarse-grained abstraction that treats per-hop retrieval and single-step reasoning as a constant factor and counts how the search over hop combinations grows. Under this abstraction, an unstructured single-agent search over the joint hop space scales as $O(s^h \cdot r^h)$, a two-stage (Plan + Execute) scheme as $O(h \cdot r + s^h)$, and OPERA's staged decomposition as $O(h \cdot s \cdot r)$. These expressions characterize orchestration-level scaling under the stated abstraction; they are not end-to-end complexity lower bounds. Each agent focuses on its specific domain while maintaining coordination through the TMC mechanism. Compared with joint training, MAPGRPO reduces cross-agent interference by optimizing each agent on its role-specific objective and induced data distribution.

\noindent \textbf{High-Score Sample Selection Analysis.} The high-score sample selection strategy reduces the variance contributed by purely policy-generated samples. Let $\sigma^2$ denote the variance of policy-generated rewards. Under pure exploration, the group-mean variance is $Var[\bar{r}_{pure}] = \frac{\sigma^2}{G}$. When one deterministic high-score reference sample is inserted into a group, only $G-1$ samples contribute stochastic variance, giving $Var[\bar{r}_{mixed}] = \frac{(G-1)\sigma^2}{G^2} = \frac{\sigma^2}{G}\left(1-\frac{1}{G}\right)$.

\noindent \textbf{Sample Complexity.} Sequential training lets each agent learn on the input distribution induced by its already-optimized predecessors, avoiding the cross-agent interaction that arises under joint optimization. Empirically (Tables~\ref{tab:combined_ablation},~\ref{tab:ood_evaluation}), this staged scheme attains higher accuracy than joint/monolithic training under a comparable data budget; we do not claim a closed-form sample-complexity separation.

\subsection{Pre-scored Dataset Construction ($\mathcal{D}_{\text{scored}}$)}

MAPGRPO uses a pre-scored dataset $\mathcal{D}_{\text{scored}}$ containing candidate samples with reward labels to address reward sparsity in early training. We describe the construction process below.

\noindent\textbf{Dataset Composition and Scale.}
$\mathcal{D}_{\text{scored}}$ contains multi-hop reasoning questions from three datasets:
\begin{itemize}
    \item \textbf{Musique (60\%):} 1,800 complex compositional reasoning questions
    \item \textbf{HotpotQA (25\%):} 750 two-hop reasoning scenarios  
    \item \textbf{2WikiMultiHopQA (15\%):} 450 questions with different reasoning patterns
\end{itemize}

\noindent\textbf{Scale and Implementation Details:} For each question, we sample multiple candidate decompositions from DeepSeek R1 offline, score every candidate, and keep the highest-scoring one, yielding a pool of 3,000 expert references. During Plan Agent training, an expert reference is intermittently injected as $c_{\text{best}}$ into the candidate group to stabilize the group-relative advantage baseline and mitigate reward sparsity; policy-gradient updates remain restricted to on-policy candidates.

\noindent\textbf{Golden Plan Standard Generation.}
For each question $q$ with ground-truth answer $a^*$ and supporting facts $facts^*$, we generate candidates using the DeepSeek R1\footnote{We use DeepSeek R1 via the official commercial API for data generation.}: 
\begin{algorithmic}[1]
\STATE Analyze ground-truth supporting facts using DeepSeek R1
\STATE Generate multiple candidate decompositions using R1's planning capabilities
\STATE Ensure placeholder dependency structure matches runtime format
\STATE Score all candidates and select the highest-scoring decomposition
\STATE Validate logical coherence of the selected decomposition through DeepSeek R1
\end{algorithmic}

\noindent\textbf{Scoring Framework.}
DeepSeek R1 is used as the external judge model for semantic planning quality. During MAPGRPO training, DeepSeek R1 also scores online rollout candidates generated by the current policy under the same planning rubric. Each candidate undergoes judge-based evaluation together with rule-based structure checking and end-to-end execution simulation. The pre-scoring function is:
\begin{align}
r_{\text{pre}}(q, c) = \sum_{i=1}^5 w_i \cdot f_i(q, c, \mathcal{E}(c)),
\end{align}
where $\mathcal{E}(c)$ represents the execution result of candidate $c$, and the scoring components are:

\begin{itemize}
    \item $f_1$: \textbf{Logical Coherence} ($w_1 = 0.25$) - Dependency validity and decomposition logic judged by DeepSeek R1
    \item $f_2$: \textbf{Execution Feasibility} ($w_2 = 0.25$) - Success and completion rates during simulation  
    \item $f_3$: \textbf{Answer Accuracy} ($w_3 = 0.30$) - Exact match with ground-truth answers
    \item $f_4$: \textbf{Efficiency} ($w_4 = 0.10$) - Plan conciseness and step optimization
    \item $f_5$: \textbf{Placeholder Correctness} ($w_5 = 0.10$) - Proper dependency modeling syntax
\end{itemize}

\textbf{Logical Coherence ($f_1$):} Evaluates dependency relationships and decomposition logic through automated analysis of sub-question ordering, placeholder usage, and logical flow consistency.

\textbf{Execution Feasibility ($f_2$):} Measures plan executability via end-to-end simulation using our retrieval pipeline, computing success rates and completion percentages across all sub-steps.

\textbf{Answer Accuracy ($f_3$):} Uses exact match scoring after answer normalization, ensuring the complete execution path leads to the correct ground-truth answer.

\textbf{Efficiency ($f_4$):} Optimizes for plan conciseness while maintaining completeness, penalizing unnecessarily complex decompositions.

\textbf{Placeholder Correctness ($f_5$):} Validates proper dependency modeling syntax and placeholder resolution mechanisms.

\noindent\textbf{Quality Control and Validation.}
We implement multiple validation layers to ensure dataset quality:
\begin{enumerate}
    \item \textbf{DeepSeek R1 Generation Quality:} Use R1's built-in reasoning verification and self-correction capabilities
    \item \textbf{Execution Simulation:} Complete end-to-end execution of each plan using our retrieval pipeline
    \item \textbf{Answer Verification:} Exact match validation against ground-truth answers with normalized string comparison
    \item \textbf{Format Compliance:} JSON structure and placeholder syntax validation using automated parsers
    \item \textbf{Diversity Filtering:} Removal of near-duplicate candidates using semantic similarity thresholds (cosine similarity $<$ 0.85)
\end{enumerate}

\noindent\textbf{Construction Algorithm.}
Algorithm \ref{alg:prescored_construction} provides the complete implementation for constructing $\mathcal{D}_{\text{scored}}$.

\begin{algorithm}[t!]
\caption{Golden Reference Construction with DeepSeek R1}
\label{alg:prescored_construction}
\small
\begin{algorithmic}[1]
\REQUIRE Mixed dataset $\mathcal{D}_{\text{mix}}$, DeepSeek R1 API service $\mathcal{R}_{\mathrm{R1}}$, retrieval system $\mathcal{R}$
\ENSURE Golden reference dataset $\mathcal{D}_{\text{scored}}$

\STATE $\mathcal{D}_{\text{scored}} \leftarrow \emptyset$
\FOR{each $(q, a^*, facts^*) \in \mathcal{D}_{\text{mix}}$}
\STATE $\{c_1,\dots,c_M\} \leftarrow \text{GenerateCandidatePlans}(q, a^*, facts^*, \mathcal{R}_{\mathrm{R1}})$ \COMMENT{$M$ candidate decompositions}
\FOR{each candidate $c_m \in \{c_1,\dots,c_M\}$}
    \STATE $\mathcal{E}(c_m) \leftarrow \text{ExecuteSimulation}(c_m, q, \mathcal{R})$
    \STATE $r_{\text{pre}}(q, c_m) \leftarrow \text{JudgeAndScore}(q, c_m, \mathcal{E}(c_m), a^*, \mathcal{R}_{\mathrm{R1}})$
\ENDFOR
\STATE $c_{\text{best}} \leftarrow \arg\max_{m} r_{\text{pre}}(q, c_m)$
\STATE $\mathcal{D}_{\text{scored}} \leftarrow \mathcal{D}_{\text{scored}} \cup \{(q, c_{\text{best}}, r_{\text{pre}}(q, c_{\text{best}}))\}$
\ENDFOR

\STATE Apply quality thresholds to reference candidates
\STATE Validate reference quality and filtering results
\RETURN $\mathcal{D}_{\text{scored}}$
\end{algorithmic}
\end{algorithm}

\subsection{Limitations and Future Directions}

OPERA shows strong multi-hop reasoning performance but has several important limitations.

\noindent\textbf{Scalability Challenges.} Performance declines substantially on questions requiring longer reasoning chains due to error accumulation and limited training data for such cases. The multi-agent architecture introduces higher computational overhead than single-pass RAG systems, which limits real-time applications.

\noindent\textbf{Retrieval Dependencies.} Even with adaptive rewriting, OPERA is limited by corpus coverage. Our analysis shows that retrieval-related issues account for a substantial portion of Musique failures. Questions requiring implicit inference or commonsense knowledge are particularly difficult for the system.

\noindent\textbf{Planning Inefficiencies.} Many questions allow multiple valid decomposition paths, but OPERA lacks explicit optimization for path efficiency and may select suboptimal reasoning chains. Several research directions could address these limitations. Developing adaptive decomposition methods with path-efficiency optimization would improve planning effectiveness. Creating specialized datasets for long-chain reasoning would enhance performance on complex multi-hop questions. Hybrid architectures that maintain modularity while reducing computational overhead could make the system more practical for real-time use. Advanced retrieval techniques like Beam Retrieval~\cite{zhang2024endtoend} could explore multiple retrieval paths simultaneously, improving robustness for ambiguous queries.

\noindent\textbf{Corpus Constraints and Domain Scope.} The primary objective of our experimental design is to validate the structural rationality of the OPERA architecture and the effectiveness of the MAPGRPO training protocol within a controlled, high-density information environment. However, a primary limitation of the current evaluation is that OPERA operates within a \textbf{closed-domain retrieval} setting. The knowledge base is synthesized from the official document collections of the three source datasets, which ensures high information density but may not fully represent the noise and ambiguity of the open web. This restricted scope limits the system's exposure to massive-scale entity disambiguation challenges. Future work will focus on extending OPERA to \textbf{open-domain scenarios} using the full Wikipedia corpus and real-time web indexing. This extension will test the architecture's robustness in navigating more diverse knowledge distributions and answering more generalized, cross-domain questions.

\subsection{Prompt Templates}
The prompt templates for each agent are presented below.

\begin{promptbox}{Plan Agent Prompt Template}
You are a strategic planning agent. Given a complex 
multi-hop question, decompose it into a sequence of 
simpler sub-goals with dependency modeling.

Question: {question}

Please generate a plan with the following JSON format:
[
  {
    "subgoal_id": 1,
    "subgoal": "First sub-question to answer",
    "dependencies": []
  },
  {
    "subgoal_id": 2, 
    "subgoal": "Second sub-question using [entity from step 1]",
    "dependencies": [1]
  }
]

Requirements:
- Use placeholder mechanism: [entity from step X] for dependencies
- Each subgoal should be answerable with a small set of documents
- Maintain logical flow and clear dependencies
\end{promptbox}

\begin{promptbox}{Analysis-Answer Agent Prompt Template}
You are an analysis and answering agent. Given a sub-question 
and retrieved documents, determine if you can answer the 
question and provide analysis.

Sub-question: {subgoal}

Retrieved Documents: {documents}

Please respond in the following JSON format:
{
  "status": "yes" or "no",
  "answer": "extracted answer if status is yes, empty if no",
  "analysis": "explain why you can/cannot answer based on 
               the provided documents"
}

Key principles:
- status="yes": Documents contain sufficient information
- status="no": Documents lack necessary information
- analysis: Always explain your reasoning
\end{promptbox}

\begin{promptbox}{Rewrite Agent Prompt Template}
You are an expert query rewriter for information retrieval.

## Rewrite Task
Original Question: {sub_goal}
Failure Reason: {failure_info}

## Current Documents Preview
{docs_preview}

## Instructions
1. Analyze why the current query failed to retrieve relevant information
2. Generate an improved search query using keyword expansion and synonyms
3. Focus on key entities, concepts, and alternative phrasings
4. Keep the rewritten query concise but comprehensive

## Output JSON Format
{
  "rewritten_query": "improved search query with expanded keywords",
  "strategy": "brief explanation of rewrite approach",
  "keywords": ["key", "terms", "and", "synonyms"]
}

Generate rewrite:
\end{promptbox}

\FloatBarrier
\end{document}